\pdfoutput=1
 \documentclass[11pt]{article}

  \usepackage{framed}         
  \usepackage{soul}           

\usepackage{subcaption}
\usepackage{comment}

\usepackage[bookmarks=false]{hyperref}
  \usepackage{rotating}
  \usepackage{floatpag}
  \rotfloatpagestyle{empty}

  \usepackage{amsthm}
  \usepackage{graphicx}

  \usepackage{blindtext}
\usepackage{graphicx}
\usepackage{multirow}
\usepackage{mathtools}
\usepackage[noadjust]{cite}
\usepackage{array}
\usepackage[labelformat=simple]{subcaption}

\usepackage[letterpaper, margin=0.63in,top= 1 in,right=1in, left=1in]{geometry}

\usepackage{cite}
\usepackage{graphicx}
\usepackage{amsfonts}
\usepackage{algorithm}
\usepackage{algpseudocode}
\usepackage{amsmath}
\usepackage[utf8x]{inputenc}

\usepackage{siunitx}

\usepackage{psfrag}
\usepackage{csquotes}

\usepackage{xspace}

\newcommand{\Xcomment}[1]{}

\newcommand*{\eg}{e.g.\@\xspace}
\newcommand*{\ie}{i.e.\@\xspace}

\makeatletter
\newcommand*{\etc}{%
    \@ifnextchar{.}%
        {etc}%
        {etc.\@\xspace}%
}
\makeatother

\algnewcommand\algorithmicforeach{\textbf{for each}}
\algdef{S}[FOR]{ForEach}[1]{\algorithmicforeach\ #1\ \algorithmicdo}

\def\LTEUploadSpeed{17.84 Mbps {}}
\def\ThreeGUploadSpeed{0.47 Mbps {}}
\def\TwoGUploadSpeed{0.03 Mbps {}}
\def\WiFiUploadSpeed{99.04 Mbps {}}
\def\EthernetUploadSpeed{93.37 Mbps {}}

\usepackage{authblk}

\begin{document}
 
 \title{Age of Information in Practice}
 
\author[*]{Elif Uysal}
\author[**]{Onur Kaya}
\author[*]{Sajjad Baghaee}
\author[***]{Hasan Burhan Beytur}

\affil[*]{Middle East Technical University }
\affil[**]{Isik University \LaTeX\ University}
\affil[***]{University of Texas at Austin}
\renewcommand\Authands{ and }
  
  \footnotetext{This work was supported by TUBITAK Grant 117E215 and by Huawei. We thank Canberk Sonmez and Egemen Sert for their work which inspired the writing of this chapter.}



  \maketitle







  
  

\begin{abstract}

While age of Information (AoI) has gained importance as a metric  characterizing the freshness of information in information-update systems and time-critical applications, most previous studies on AoI have been theoretical. 
In this chapter, we compile a set of recent works reporting AoI measurements in real-life networks and experimental testbeds, and investigating practical issues such as synchronization, the role of various transport layer protocols, congestion control mechanisms, application of machine learning for adaptation to network conditions, and device related bottlenecks such as limited processing power. 

\end{abstract}

\section{Introduction}
\label{introduction}
Theoretical studies have by now  established Age of Information (AoI) as a Key Performance Indicator (KPI) that characterizes information freshness in status update systems and applications that are time-critical. Several principles of age optimization have been developed for network models under simplifying assumptions, when delay and service time statistics are known. However, in practice, it may be difficult, if at all possible, to extract these statistics and optimize for them in real life network implementations due to complex interactions among different networking layers. Moreover, there are a number of practical system issues that have been largely ignored in theoretical problem formulations exploring the fundamental behaviour of age in networks.

There have been a relatively small number of implementation studies on AoI to date \cite{core-aoi,canberk2018,EgemenRL,Shreedhar2019,8806423,9049792,Kadota2020Poster, kadota2020wifresh,Kaul2019Vehicular}.
The goal of this chapter is to compile and discuss some results on practical aspects of AoI from this burgeoning literature. We will review experimental and emulation based AoI measurement results for data flows that are transmitted over wireless/wired links, using different transport layer protocols. Practical issues such as synchronization and the suitability of various congestion control mechanisms and application layer approaches to control AoI will be covered. The impact of device related bottlenecks on age, such as limited processing power or age constraints that may affect simple IoT nodes will be demonstrated.

The first emulation study of AoI in wireless links was reported in \cite{core-aoi}, and the first real-life implementation measuring the variation of AoI over TCP/IP links served by WiFi, LTE, 3G, 2G and Ethernet was demonstrated in \cite{canberk2018}. The experimental AoI measurement results reported in \cite{canberk2018}, exhibit a non-monotone (specifically, \enquote{U-shaped}) AoI vs arrival rate characteristic for TCP/IP connections served by Wifi, Ethernet and LTE links. This measured charecteristic is in line with theoretical results for FCFS systems with Poisson or Gamma distributed arrivals \cite{gammaAwakening}.   

According to the U-shaped relation between AoI and sampling rate in the real-life TCP/IP connection in \cite{canberk2018}, the age first falls sharply as the update arrival rate increases. Then, it stays relatively flat. After a certain arrival rate, though, due to building congestion in queues, a sharp increase in age is observed. This indicates that there is an acute need for age-optimal service policies in practical networks, for several reasons: Firstly, the increasing importance of machine type communications including autonomous systems, remote monitoring and control, implies an increase in the type of services where freshness of data, rather than throughput, is the main performance criterion. Unlike conventional data transmission, where the goal is to transmit the entirety of a stream of data as reliably as possible, in status-update type services only the timely data matter. Increasing the rate at which samples are injected into the network starts acting contrary to the purpose of freshness after a certain point. Secondly, the goals of achieving low AoI and high throughput are often aligned, as we shall see in the system implementation examples in the following sections. The principal reason behind this is that a good AoI performance requires a high throughput of data updates, at sufficiently low delay. 

Therefore, we believe that AoI optimization should be one of the directions for the future development of network architectures. By now, there is a number of years of theoretical work supporting this vision, and in the rest of this chapter we will elaborate on the initial implementation experience in various layers of the network protocol stack.

One of the challenges in translating theoretical work on the control of AoI to practically viable algorithms is that theoretical work often assumed full knowledge of the statistics of network delay (e.g. \cite{UpdateOrWait, ClementRL}).
Service or age-control policies proposed in the literature rely heavily on the delay statistics of the network. However knowledge of these statistics is not easy to obtain, especially as the scale of the network, and hence, the number of connections grow. Basing a decision policy on ill-inferred or invalid statistics can be a suboptimal approach. In such scenarios, AoI aware sampling, scheduling and transmission policies can be adaptively generated using machine learning.  The idea of introducing a delay to the sampling operation in response to instantaneous age and the delay statistics on the network was proposed in \cite{UpdateOrWait}. However, this smart sampling operation was based on a knowledge of the statistics of the network delay. In \cite{EgemenRL,ClementRL},  reinforcement learning was employed to obtain this sampling operation without requiring prior information about the delay statistics. Similarly, in \cite{8685524,Ceran_INFOCOM2019,Abd-Elmagid2020IEEETrans,Abd-Elmagid2020IEEEVehTrans} age-aware scheduling is studied using RL methods. 

The rest of this chapter is organized as follows: Section \ref{aoi-intro} provides the definition of AoI, expressed in terms of time-stamps of transmitted and received packets, followed by a brief review of the  behaviour of steady-state average AoI under some basic queuing and service disciplines, in order to provide a basis for the interpretation of practical results in the following sections. Section \ref{physical-testbed} describes the measurement of AoI on a simple physical network testbed. Section \ref{AoI_computing_issue_realistic_setups} discusses general issues arising while computing age in realistic setups, such as clock bias, and how to mitigate those. Section \ref{sec:TCP} reviews AoI in TCP connections, first in an emulation testbed in subsection \ref{emu-testbed}, and then a real-world testbed testing TCP/IP connections running on various physical links such as WiFi, Ethernet, LTE, 3G  and  2G  in  subsection  \ref{Real-world-testbed}. Section \ref{AoI_UDPTCP} contrasts age behavior over UDP and TCP connections. Section \ref{AoI_MachineLearning} overviews the application of statistical learning methods to age optimization in practical networks. Section \ref{sec:ACP} reviews recent proposals of application-layer mechanisms for age control over UDP, in particular the ACP protocol.
Section \ref{wireless_access_scheduler} describes implementation of age-aware scheduling in Wi-Fi uplink or downlink settings.  Section \ref{conclusion} concludes the chapter by offering a general vision for the future development of this research area.

\section{Age of Information: definition, measurement and behaviour in queuing systems}
\label{aoi-intro}

\begin{figure}
\centering
	\includegraphics[width=1\linewidth]{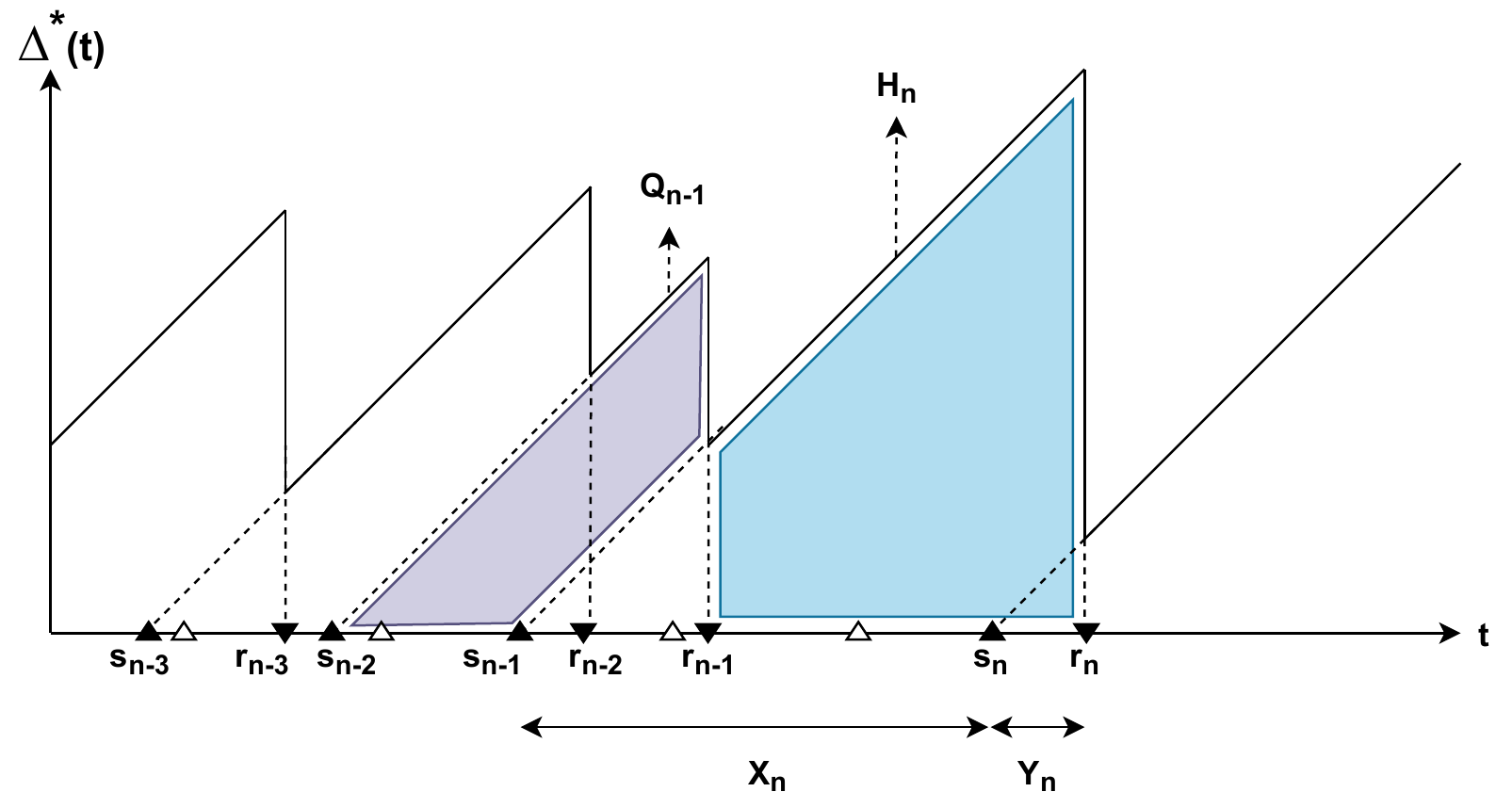}
	\caption[Age sample path $\Delta(t)$ for a given realization of packet transmission and reception times $r_i$ and $s_i$.]{Age sample path $\Delta(t)$ for a given realization of packet transmission and reception times $r_i$ and $s_i$. \copyright [2020] IEEE. Reprinted, with permission, from \cite{9049792}.}
	\label{aoi1}
\end{figure}

Status age of a flow is defined as $\Delta(t) = t - U(t)$, where $U(t)$ is the generation time (i.e. time stamp) of the newest data packet belonging to this flow that has been received by the destination by time \mbox{time $t$}. As a consequence of this linearity, $\Delta(t)$ follows a saw-tooth pattern as seen in Fig. \ref{aoi1}. In between updates, age increases linearly in time and drops just after a new update is received. 

As shown in Fig. \ref{aoi1}, the area under age sample paths can be presented as sum of the trapezoidal areas $\{Q_i\}$, or alternatively, that of the areas $\{H_i\}$.

Let $n(T)$ be the number of received packets by time $T$. In what follows, we will use $n$ and $n(T)$ interchangeably, depending on whether or not the time dependence needs be explicitly stated in an expression. The area is composed of the area of polygon $Q_1$, isosceles trapezoids $Q_i$'s for $2\leq i \leq n$ and the triangle of base $Y_n$ .

\begin{equation}
\overline{\Delta} =  \lim_{T\to\infty}\frac{Q_1 + \sum_{i=2}^{n(T)} Q_i + Y_{n(T)}^2/2}{T}
\label{eq:age with q}
\end{equation}

As ${T\to\infty}$, it also follows that ${n(T)\to\infty}$, and that the average age is determined by the total area of the isosceles trapezoids normalized by time. This area depends only on the arrival and departure instants.

\begin{equation}
Q_i = \frac{(2r_i - s_i - s_{i-1})(s_i - s_{i-1})}{2}
\label{eq:qs}
\end{equation}

Note that, $(s_i - s_{i-1})=X_i$, the inter-arrival time between successfully transmitted packets.

 The formulation using $Q_i$'s facilitates mathematical analysis, however, a formulation using $H_i$'s is useful in some scenarios involving practical measurement of AoI. Note that:

\begin{equation}
\overline{\Delta} = \lim_{T\to\infty}\frac{1}{T} \sum_{i=1}^{n(T)} H_i
\label{eq:age with h}
\end{equation}

\begin{equation}
H_i = (r_i - r_{i-1})(r_{i-1}-s_{i-1}) + \frac{(r_i - r_{i-1})^2}{2}
\label{eq:hs}
\end{equation}

where $(r_i - r_{i-1})$ is the inter-departure time between $i-1^{th}$ and $i^{th}$ packets and $(r_{i-1}-s_{i-1})$ is equal to $Y_{i-1}$, which is the system time of $i-1^{th}$ packet.

Depending on the location of the age or age penalty computation, it may be more appropriate to measure the age at the receiver or at the transmitter. The location of this computation in scheduling/transmission algorithms that control age often depends on the capabilities of the devices or the essence of the control problem. For example, age of a flow between a simple transmitter and a remote server should be measured at the destination (i.e. the server). However, if the transmitter needs to control the AoI, then it would need to know the AoI values with minimum latency. In such a situation, the AoI is measured at the transmitter. 

Note that, as seen in (\ref{eq:qs}) and (\ref{eq:hs}), the timestamps $r_i$'s and $s_i$'s should be computed using a common time reference. In practice, however, as the receiver and transmitter are separate machines, they rarely have access to a common clock, which results in a synchronization issue that we will address in Section \ref{synchronization}. 

Peak age of information is another metric often investigated in literature:

\begin{equation}
    \overline{\Delta_{peak}} = \frac{1}{n} \sum_{i=1}^n\Delta(r_i^-)
\end{equation}

This quantity can also be calculated using the transmission and reception instances of packets

\begin{equation}
    \overline{\Delta_{peak}} = \frac{1}{n} \sum_{i=1}^n (r_i - s_{i-1})
    \label{eq:paoi}
\end{equation}

\begin{figure}[ht]
\centering
    \begin{psfrags}
    \psfrag{m}[l]{\tiny{M/M/1}}
    \psfrag{n}[l]{\tiny{M/D/1}}
    \psfrag{k}[l]{\tiny{D/M/1}}
    \psfrag{l}[t]{\small{Load, $\rho$}}
        \psfrag{a}[bl]{\small{avg. Age, $\bar{\Delta}(t)$}}
        \psfrag{d}[bl]{\small{avg. delay}}
    \includegraphics[width=0.8\linewidth]{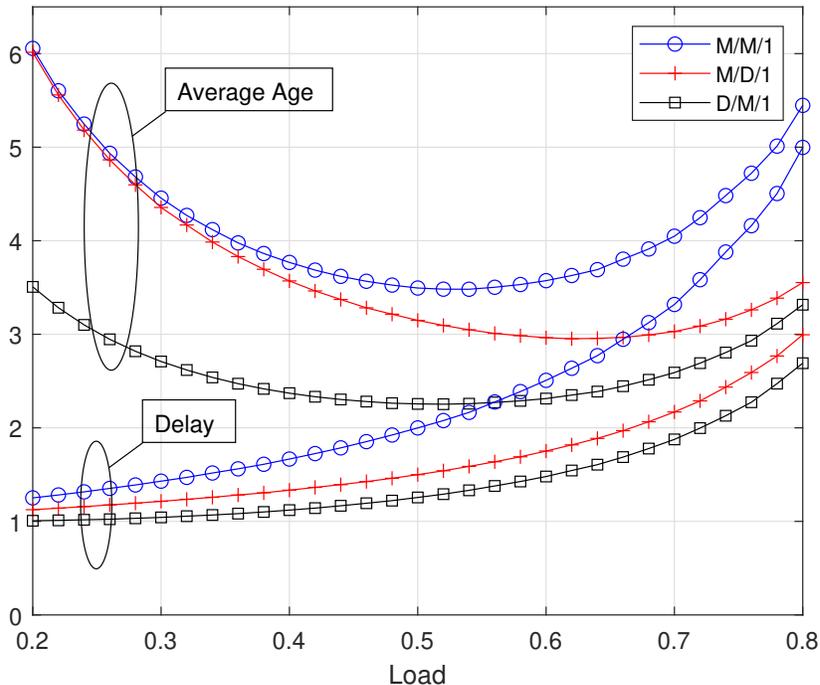}
    \end{psfrags}
    \caption[Avg. delay and AoI attained by an M/M/1 queuing systen with service rate $\mu=1$.]{Avg. delay and AoI attained by an M/M/1 queuing systen with service rate $\mu=1$. \copyright [2018] IEEE. Reprinted, with permission, from \cite{canberk2018}.}
\label{fig:AgevsDelay}
\end{figure}

It should be stressed that minimizing status age is not the same as minimizing delay. It requires, on the other hand, a joint optimization of delay and throughput together \cite{KaulYatesGruteser-Infocom2012}. We exhibit this phenomenon on simple queuing systems in Fig. \ref{fig:AgevsDelay}, with Poisson packet arrivals. We plot average delay and AoI as a function of load to the server (the server corresponding to the communication channel in this case). The transmission duration of each packet on the link is modeled as an exponential random variable, independent of all other transmission durations. The load is equivalent to throughput, as long as the queue is operating in the stable regime. This abstraction is an M/M/1 queuing system\footnote{An M/M/1 queue is a single-server queuing system abstraction. Packet inter-arrival times are exponentially distributed and i.i.d., with mean $1/\lambda$. Service times are i.i.d. exponential with mean $1/\mu$.  The expected delay experienced by packets at steady state is $(1/\mu)/(1-\rho)$, where $\rho=\lambda/\mu$ is the \enquote{load}, i.e., the steady-state probability that the system is busy. Steady-state delay therefore monotonically increases with the load and blows up as $\rho \to 1$, and the system enters an unstable regime (where delay is not guaranteed to be finite) when $\rho\geq 1$. If discarding packets from the buffer is allowed (as in an M/M/1/k system that holds up to $k-1$ packets in queue), the delay will remain finite at the expense of dropping a fraction $\max(0,\rho-1)$ of all arriving packets.} with a first-come first-served (FCFS) discipline. 

The throughput increases linearly with load, $\rho$, in the range of values plotted, as the system is stable in this range. Note that the AoI \textbf{decreases} with increasing load up to the age optimal operating point $\rho*=0.53$, as arriving packets get served with no appreciable waiting time in queue- as the queue is often empty up to that point of optimal server utilization. When packets get served with nearly no delay, they contribute to a reduction of age upon service completion. The average number of packets in the system at $\rho=\rho*$, is $\approx 1.13$. However, AoI starts increasing due to queuing effects as the load exceeds $\rho*$: this is a natural consequence of the FCFS queuing discipline which lets newer packets get stale waiting in queue behind older packets. Interestingly, we will see in the experimental results on TCP connections in the Internet in the following sections that the optimality of a load value around $0.53$, and a total number of packets in flight being around $1$ continue to hold, even though the underlying arrival and service processes are much more complex.

Note from the plots in Fig. \ref{fig:AgevsDelay} AoI exhibits different behavior than delay. Increasing update rate, while it may contribute to a larger throughput of updates, does not necessarily improve AoI. In fact, it can even result in an increase in AoI in the FCFS case. Better AoI is achieved in multi-server models (\eg, M/M/2
\cite{KamKompellaEphremidesTIT}, M/M/$\infty$) due to the increased service rate: the AoI in an M/M/2 system is nearly half of that in M/M/1. Note, though, that increasing the update rate in such systems may cause some packets to become obsolete by the time they arrive at the destination, because they have been preceded by newer packets. This makes packet management schemes important~\cite{CostaCodreanuEphremides_TIT}. The fraction of packets that are rendered obsolete, causing a waste of resources without a reduction in age, increases with the load $\rho$. Mitigating this problem has been the subject of previous studies (\eg, \cite{CostaCodreanuEphremides_TIT}) which studied packet management schemes.

The literature on AoI analysis and optimization (\eg, \cite{KaulYatesGruteser-Infocom2012,2012CISS-KaulYatesGruteser,2012ISIT-YatesKaul,Kam-status-ISIT2013, KamKompellaEphremides2014ISIT,CostaCodreanuEphremides2014ISIT,2015ISITHuangModiano, CostaCodreanuEphremides_TIT}) was based on exogenous packet arrivals, prior to ~\cite{7308962}. The "generate-at-will" model was introduced in \cite{7308962, YinINFOCOM2016, SunJournal2016}, where a source can generate and output packets to deliberately control age. The model in \cite{SunJournal2016} will be important as we consider arrival rate control in the application later, on top of various transport layer mechanisms, to obtain favorable age performance in the presence of network bottlenecks. Hence, below we elaborate on that model.

In \cite{SunJournal2016}, updates $i=1, 2,\dots$ generated by a source were subject to a stochastic process of delays $\{Y_i\}$ in the network. The delay process was arbitrary and not affected by the actions of the transmitter. This modeled a scenario where the packet flow is injected to a large network where the congestion and delays is caused by a multitude of factors, among which the actions of this particular source is negligible. The source is assumed to get instantaneous feedback when an update is delivered, which corresponds to neglecting the delivery time for acknowledgements.  The policy for generating updates is determined by the sequence of waiting times $\{Z_i\}$. A possible work-conserving policy is the zero wait (ZW) policy, which achieves maximum link utility and minimum average delay by setting $Z_i=0$ for all i. This corresponds to submitting a fresh update once the server (\ie, communication link) becomes idle.  

\cite{YinINFOCOM2016} investigated the performance of the ZW policy with respect to AoI metric, based on the assumption that it is feasible. Somewhat counter-intuitively, ZW is sub-optimal for a large class of delay distributions. The results in \cite{YinINFOCOM2016,SunJournal2016} characterized policies that minimize not only AoI, but a general AoI penalty function $g(\Delta(t))$. Often, an optimal update policy sends updates at a rate that is lower than the maximum at which it is allowed to send. This pointed to a very important conclusion: one could control AoI by judiciously adapting the rate of generation of samples (updates) to variable external conditions such as delay. This can be done through the measured or estimated \cite{7308962} value of the age penalty. Such policies can prevent unnecessary staleness resulting from congestion, queuing, or more generally, inefficient utilization of resources from an overall network perspective.

In the rest of the chapter, we will explore how the conclusions drawn from the analysis of idealized queuing and service models and age control policies from literature extend to real-life networks. 

\section{AoI measurement on a simple physical network}
\label{physical-testbed}

Age measurement in a real network requires, first and foremost, an accurate timing information at the node where the calculation will be carried out. As the clocks at the client and server nodes are not necessarily synchronized, either an estimation of the timing offset, or a system-wide synchronization must be performed prior to the computation of age. 

Before discussing synchronization methods in relation to age measurements and the effect of eventual synchronization/timing estimation error on age performance, we first present a toy network model from \cite{8806423} which circumvents such need for synchronization, so that the 
effect of sampling rate on A
age can be observed over a network running a practical transport layer protocol. 

\begin{figure}[ht]
	\centering
	\includegraphics[width=0.99\linewidth]{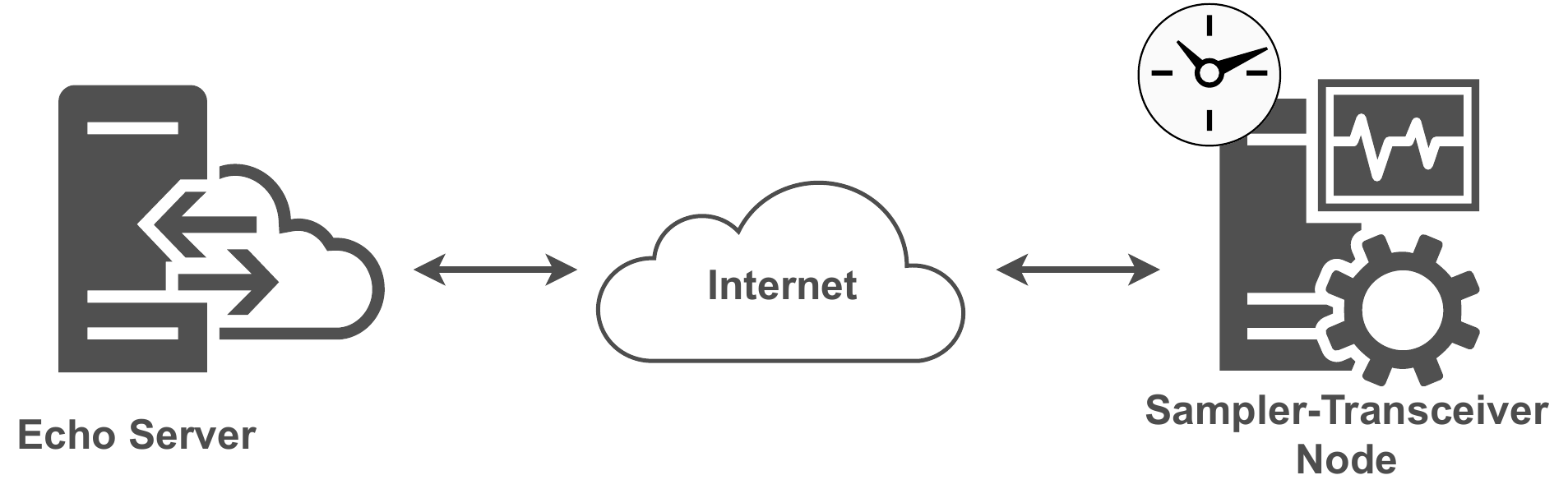}
	\caption[An illustration of the physical testbed.]{An illustration of the physical testbed. \copyright [2019] IEEE. Reprinted, with permission, from \cite{8806423}. }
	\label{testbed}
\end{figure}

The model comprises a sampler-transceiver node in Ankara, Turkey, and an echo server (Fig. \ref{testbed}) Istanbul, Turkey. The sampler-transceiver  produces and sends UDP packets which contain the timestamp marking the generation time, packet ID and a dummy payload and transmits the packet to the echo server through the Internet. The echo server echoes the received packet immediately after receiving it. Upon reception and decoding, the receive timestamp of each packet is noted and compared with generation timestamp, and the instantaneous age is calculated. A multi-threaded or multi-process solution, which enables simultaneous sampling, transmission and reception packets is employed at the  sampler-transceiver. 

\begin{figure}
	\centering
	\includegraphics[width=0.9\linewidth]{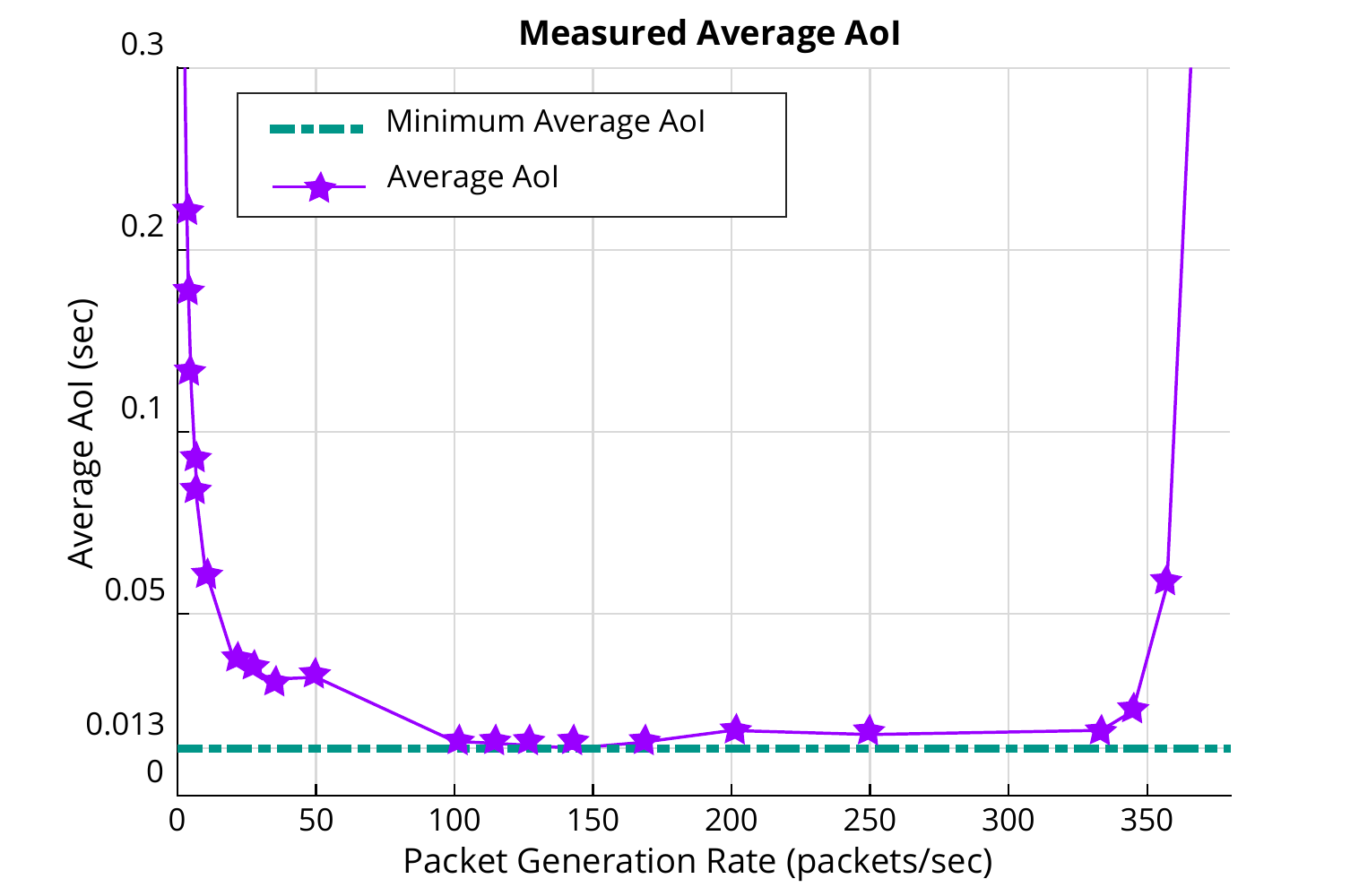}
	\caption[AoI vs sampling rate, on the testbed running UDP protocol.]{AoI vs sampling rate, on the testbed running UDP protocol. \copyright [2019] IEEE. Reprinted, with permission, from \cite{8806423}}
	\label{fig:UDPAoIMinAoIresult}
\end{figure}

The two-way connection between the sampler-transceiver node and the echo Server can alternatively be viewed as a single end-to-end connection from the transmitter node to the receiver node through the echo server which may simply be seen as one of intermediate nodes in the network. Since the transmitter and the receiver are the same device, 
the average AoI can be measured without any synchronization issue. To do so, UDP packets consisting of 1058 bytes (including header) are transmitted at a constant rate. The average AoI is calculated by plugging the timing information available at the transceiver in \eqref{eq:age with h}. The sampling rate is linearly increased, from 1 packet per second up to 370 packets per second, to investigate the age-rate relation, which is plotted in Fig. \ref{fig:UDPAoIMinAoIresult}, 

Even though many routers and switches in the Internet infrastructure use FCFS queues, still, it is rather surprising to observe similar results to those obtained in the theoretical analysis of a simple M/M/1 queuing system  \cite{6195689}. Another interesting result is that, unlike the results for M/M/1, M/D/1 and D/M/1 queues shown in Fig. \ref{fig:AgevsDelay}, the average AoI is much less sensitive to sampling rate, over a large range of rate values. For example, the average age remains almost constant as the packet generation rate is increased from 100 pps up to 340 pps. This phenomenon can be explained by (i) the determinism in service time on individual links, (ii) the load being shared by multiple servers due to data being routed on multiple possible paths. It should be noted, however, that packet drops sharply increase as the rate at which samples are injected into the network increases beyond a point. 

The experimental results in Fig \ref{fig:UDPAoIMinAoIresult} also verify the expected relation between the minimum average age and the round-trip-time experienced by the packets in the network. In the testbed, before starting transmission, test packets were used to estimate the number of hops, and the average RTT from the transmitter to receiver. It was observed that each packet travels through a total of 18-20 hops from the transmitter to echo server and back, with an average RTT of 12.5 msec. Since, during the transmission of the test packets, the network was not loaded, the average RTT measured was close to the minimum possible RTT, which is essentially the sum of the round trip propagation delay and the total transmission time for one packet along its route. For a deterministic sequence of packet transmissions at rate $R_p$, the minimum average age of a packet received at the transceiver will be,
\begin{align}
    \Delta_{\min}=RTT+\frac{1}{2R_p}
    \label{min_age}
\end{align}
in the absence of queueing delay. The real life measurements are in close agreement with this estimate: the age is minimized at $13$ msec, around $R_p=140$ Hz. At this rate, \eqref{min_age} predicts an average age estimate of about 16 msec computed using the average RTT, which is expected to be slightly conservative, due to the actual minimum RTT being shorter.  

The synchronization, which was automatic due to the toy network model used in this section which involves an echo server is not possible in real life networks where the transmitter and the receiver are not co-located. We will present practical AoI measurement results for such networks in the remainder of this chapter. In order to do so, we need to first discuss possible methods of AoI measurement, synchronization, and effect of synchronization errors on AoI optimization.

\section{Clock bias and synchronization for age measurement over networks}
\label{AoI_computing_issue_realistic_setups}
Depending on the application and the network model, the AoI and other AoI-related values may be calculated at the receiver, the transmitter, or centrally. In this section we will discuss some practical timing related issues regarding AoI measurement.

\subsection{Age Measurement via Synchronization}
\label{synchronization}
A straightforward approach to AoI measurement is to establish synchronization among the transmitter and receiver clocks. This may be realized in a variety of ways, some of which are listed below:

\begin{itemize} 
\item {\textbf{Using a GPS clock}}: Using a GPS module is currently one of the most accurate methods for synchronization of distant machines. According to \cite{GPSCLOCK}, the accuracy of the GPS clock is in the order of tens of nanoseconds. GPS clocks are currently employed in UAVs, base stations, some software defined radios \cite{kadota2020wifresh}. However, this method may be unsuited for networks composed of many low power devices such as sensor modules and IoT nodes, where it is not feasible to support GPS modules on each node.

\item {\textbf{Pre-synchronization using a common reference}}: For some applications, it is possible to physically connect two devices before deploying them in separate locations. In such a case, the receiver and transmitter can be synchronized using a common Real Time Clock (RTC). The disadvantage of this approach is that the synchronization will be lost if one of the devices is reset or powered off. 

\item {\textbf{Synchronization using Network Time Protocol (NTP)}}:
NTP is used to synchronize the clocks of all participating computers in a network. It provides  synchronization in the range of 0.1 milliseconds in fast LANs and within the range of a few tens of milliseconds in the intercontinental Internet \cite{NTP}. The performance of the NTP is effected by jitter on the network, which reduces accuracy.

\item {\textbf{Synchronization via prior signalling:}}
For purposes of AoI measurement, since only the timing information at the transmitter and the ultimate receiver are required, the synchronization can be directly performed at the application layer by measuring at the transmitter the round trip time based on a simple ping mechanism prior to transmission. The receiver responds to the ping by sending the time-stamp of the received packet, and the transmitter can correct its clock assuming that the round trip delay is twice the unidirectional delay in either direction. This assumption is justifiable on the basis that the packets used in this phase are small, hence incurring negligible transmission times in queues, and the RTT times can be thought of as being dominantly due to delays. 
\end{itemize}

Once the synchronization is established, the AoI measurement can take place at the transmitter or the receiver:

\begin{itemize}
\item {\textbf {Measuring AoI at Receiver:}} To measure the AoI at the receiver, the transmitter needs to send the generation timestamps of packets. Both UDP and TCP headers carry the timestamp information of the packet, so these can be used in AoI measurement. If the header of the packet is not available, then the timestamp information can be inserted in the payload. For the calculation average AoI, (\ref{eq:age with h}) is preferable, because of its higher accuracy while averaging in short time-windows.
    
\item {\textbf {Measuring AoI at Transmitter:}} To measure AoI at the transmitter, the received time-stamps of the from the receiver are needed to be fed back to the receiver. This information can be carried using an ACK for each packet, by inserting it into the payload of the ACKs. The drawbacks of this method are the additional delay in calculation due to need for the extra wait until the ACK is received, and miscalculations that may be caused by ACKs being lost in the return link. Finally, in cases where ACK is not already used as part of the protocol and is trasmitted solely for AoI measurement, this method occupies more channel resources than measurement at the receiver. However, in setups where we need to adaptively control the age and related parameters by adjusting the transmission policy, measuring the age at the transmitter is a more viable option. 
\end{itemize}

\subsection{Effect of imperfect synchronization on age related parameters}

The methods highlighted in the previous section have the drawback that they are all bound to yield some synchronization error, which in turn translates directly to error in age measurements. In this subsection, the effect of the synchronization error on age computation is investigated.

To calculate AoI-related values, such as average peak AoI or average AoI, time-stamps from both the receiver and the transmitter are required. However, since the receiver and transmitter are distant from each other, they have their own system clocks, and even after synchronization there will be constant bias, due to the synchronization error. Assuming the drift of each clock is negligible during the time window of interest, the bias between two clocks will remain the same.

Let us denote by $s_i'$ and $s_i$, the time stamps marking the transmission of the $i$th packet, at the receiver and the transmitter respectively. That is, the transmitter knows the actual transmission time, and the receiver knows it up to a synchronization error, i.e., 

\begin{equation}
    s_i' = s_i + B
\label{eq:clockbias}
\end{equation}
for some constant bias (synchronization error) $B$. 

Similarly, let $r_i'$ and $r_i$ denote the time stamps which mark the instant of reception of the packet at the receiver and the transmitter respectively. We have,
\begin{equation}
    r_i' = r_i + B
\label{eq:r_i'}
\end{equation}

The receiver returns its perceived receive time stamp $r_i'$ to the transmitter, and the transmitter then calculates the average AoI using Eq. \eqref{eq:age with q}.  
Substituting \eqref{eq:r_i'} in \eqref{eq:qs} and \eqref{eq:hs}, we get
the erroneous areas of the trapezoids

\begin{equation}
Q_i' = Q_i + B(s_i - s_{i-1})
\end{equation}
and
\begin{equation}
H_i' = H_i + B(r_i - r_{i-1})
\label{eq:HQ biased}
\end{equation}
Substituting \eqref{eq:HQ biased} in \eqref{eq:age with h}, the average age becomes 
\begin{equation}
    \overline{\Delta}^{'} = \frac{1}{T} \sum_{i=1}^{n(T)} (H_i + B(r_i - r_{i-1}))
    \label{eq: average age with T}.
\end{equation}
Finally, since the total time elapsed, $T$, is equal to the sum of interarrival times at the receiver, $\sum_{i=1}^{n(T)} (r_i - r_{i-1})$, from \eqref{eq: average age with T}, the average age measurement in the presence of synchronization error yields
\begin{equation}
    \overline{\Delta}^{'} = \overline{\Delta} + B.
    \label{eq: age with bias}
\end{equation}
That is, the average age is shifted due to synchronization error by the same bias among the clocks of the transmitter and receiver.
Similarly, using \eqref{eq:paoi}, the time synchronization error adds the same constant bias on average peak AoI.

\begin{equation}
     \overline{\Delta_{peak}}^{'} =  \overline{\Delta_{peak}} + B
\end{equation}

At this point, an important observation is in order: the effect of error in synchronization, or equivalently, a constant bias among the transmitter and receiver clocks, simply shifts the average AoI and average peak AoI on age axis by the same bias. Therefore, even if there is an unknown bias among clocks is not known, the optimal operating point on the AoI vs sampling rate curve remains unchanged. This opens up the possibility of skipping the synchronization before age calculation altogether, in case the goal is to adjust the tranmission rate, provided the age can somehow be estimated. This idea will be revisited in Section \ref{RTTestimate}.

While a constant bias in the clock measurements has no effect on the value of the optimal sampling rate when the penalty function is a linear function of age, there may be penalty functions that are nonlinear functions of age ~\cite{YinINFOCOM2016,SunJournal2016}, for which the optimal operating points differ for biased and unbiased age. For an age penalty function, $f(\Delta(t))$, let $F(t)=\int_0^tf(\tau)d\tau$. Then, the time average age penalty is \cite{9049792}

\begin{multline}
    \overline{\Delta_{Bias}} = \frac{1}{T} \sum_{i=1}^{n(T)} F(r_i+B-s_{i-1}) - F(r_{i-1}+B-s_{i-1}) \\
     - F(r_i-s_{i-1}) + F(r_{i-1}-s_{i-1}).
     \label{eq:agebias}
\end{multline}


The age biases for the commonly used  linear ($f(t) = \alpha t$), exponential ($f(t) = e^{\alpha t} -1$) and logarithmic ($f(t) = \log(\alpha t +1)$) penalty functions \cite{NonlinearAge_Kosta,Klugel2019}, calculated using \eqref{eq:agebias} are \cite{9049792},




\begin{equation}
    \overline{\Delta_{Bias, Linear}} = \alpha B;
    \label{eq:linear}
\end{equation}{}

\begin{multline}
     \overline{\Delta_{Bias, Exp}} = \frac{1}{\alpha \sum_{i=1}^{n(T)} (r_i - r_{i-1})} \Big( e^{\alpha (\theta + B)} -
     e^{\alpha (\beta + B)} - e^{\alpha \theta} + e^{\alpha \beta} \Big)
\end{multline}
and
\begin{multline}
     \overline{\Delta_{Bias, Log}} = \frac{1}{\sum_{i=1}^{n(T)} (r_i - r_{i-1})} \bigg( \frac{1}{\alpha} \Big( \log(\alpha \beta+1) - \\ \log(\alpha \theta +1) +\log(\alpha (B +\theta)+1) - \log(\alpha (B +\beta)+1) \Big) - \\ \theta \log(\alpha \theta +1) +\beta \log(\alpha \beta +1) + \log(\alpha (B + \theta)+1)(B + \theta) \\
     -\log(\alpha (B + \beta)+1)(B + \beta)\bigg),
\end{multline}
respectively; where $\beta = r_{i-1} - s_{i-1}$ and $\theta = r_i - s_{i-1}$.



Since sampling, remote control and tracking formulations are known to benefit from non-linear age penalty functions \cite{NonlinearAge_Kosta,Klugel2019,8764465}, the measurement of correct age to be used in such penalty functions is crucial. Therefore, synchronization is essential in applications requiring non-linear penalty functions. For cases when a linear relation between age penalty and measured age exists, the bias in age measurement can be tolerated, and the following method for AoI estimation can be used in lieu of synchronization.

\subsection{Asynchronous Estimation of AoI using RTT}
\label{RTTestimate}
In cases when the exact value of age is not crucial, the transmitter, upon receiving an ACK from the receiver in response to its transmission of a packet, can directly use the RTT of that packet as an estimate for the age of that packet  \cite{Shreedhar2019}. Note that, this method doesn't require the transmitter and receiver to be synchronized. The idea is that, since the ACK doesn't carry timestamp information, it is shorter. Consequently, the error probabilities, and delays faced by the ACKs are lower, and the round trip time can be mostly attributed to the forward link. This method clearly provides an overestimate of average AoI. Yet, it is very suitable for scenarios when the age penalty function is linear, and the bias in age measurement therefore does not affect the optimal operating point. We will review some experimental results that use this method for age estimation in Section \ref{AoI_UDPTCP}.

\section{The effect of the access network to AoI in TCP-IP connections}
\label{sec:TCP}

We begin by reviewing the results of \cite{canberk2018}, which was a first attempt at studying age behavior on experimental implementations of packet flows in the Internet. 
The experimental setup was used to measure the relation between AoI and status update generation rates, for an end-to-end TCP flow traversing a physical network. The first hop on the network was varied between WiFi, Ethernet, LTE, 3G and 2G access links. Additionally, an emulation study was carried out in order to understand the effect of bottlenecks introduced at various links on the relationship between the sample generation rate and end to end AoI.

This experimental study was mainly motivated by the following question: is there a non-monotonic convex relationship between AoI and sampling rate in practice, as the one in simple idealized networking setups (e.g., Fig. \ref{fig:AgevsDelay})? This is a valid question for the following reason: practical communication networks consist of many interacting protocol components at different layers, each of which have their own queuing mechanisms. These tend to be beyond the control of the end-to-end application. In the rest, we elaborate on the results of this study that paved a way for the rest of the works we consider in this chapter.

\subsection{AoI measurement over an emulation testbed}
\label{emu-testbed}

The emulation component of the work in \cite{canberk2018} used an open-source network emulator called CORE \cite{core-intro} which was also employed in \cite{core-aoi}.

\begin{figure}
\begin{subfigure}{.5\textwidth}
  \centering
  \vspace{0.25 in}
	\includegraphics[width=0.9\linewidth]{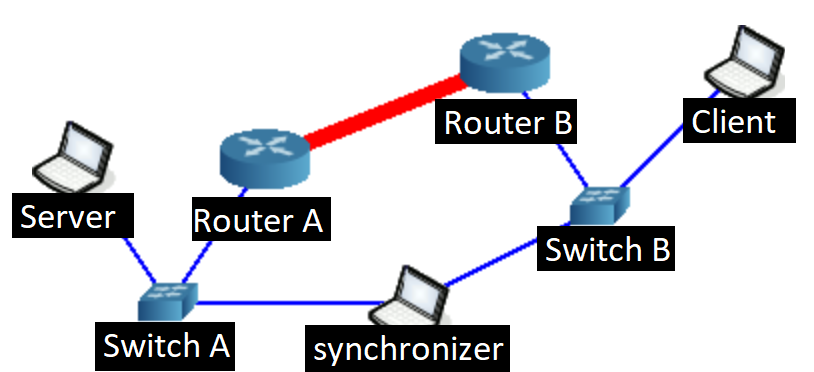}
	\vspace{0.4 in}
	\caption[CORE networking topology.]{CORE networking topology.}  \copyright [2018] IEEE. Reprinted, with permission, from \cite{EgemenRL}
	\label{core:netwtopo}
\end{subfigure}%
\begin{subfigure}{.5\textwidth}
  \centering
  \includegraphics[width=0.9\linewidth]{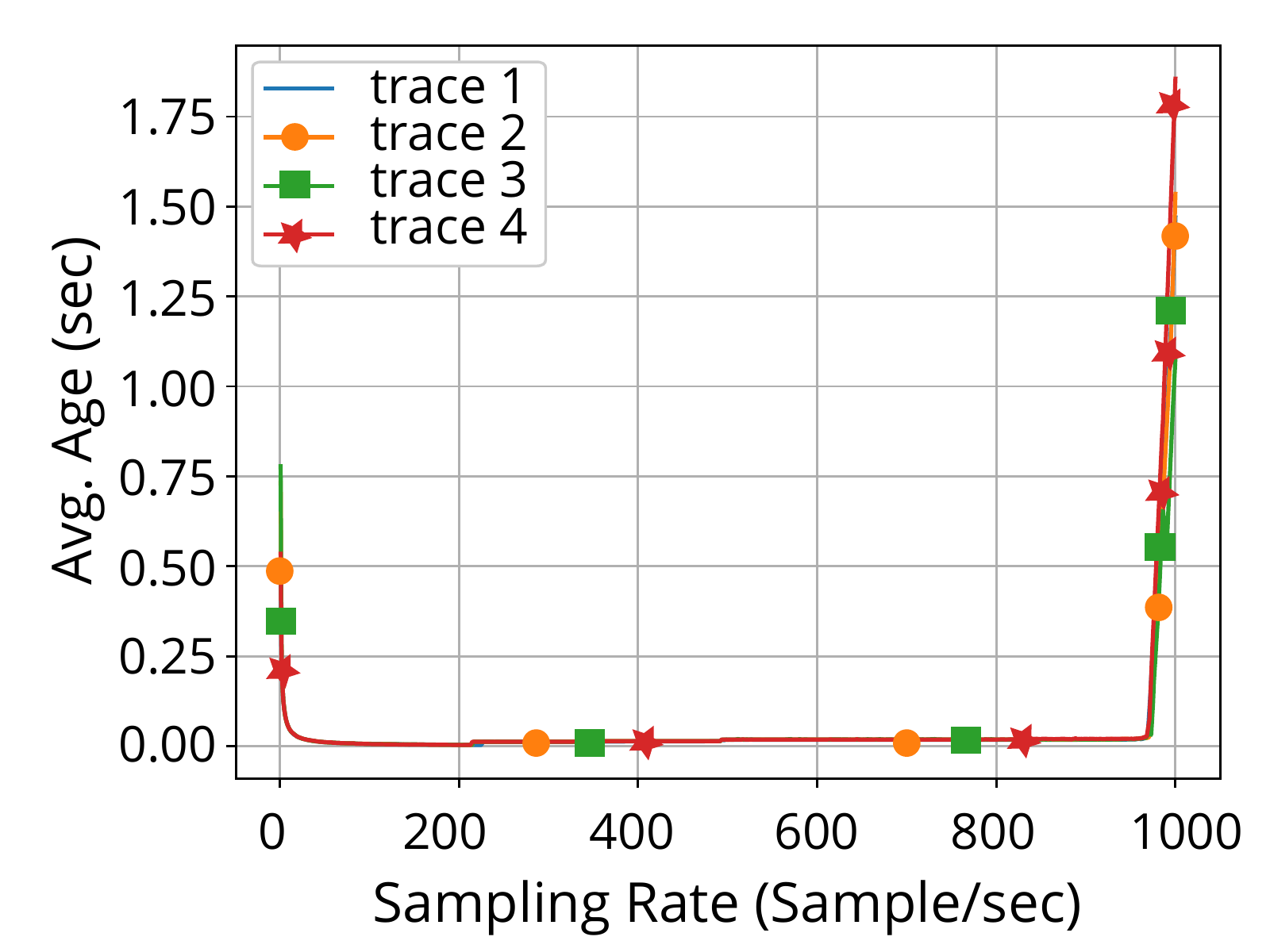}
  \caption[Age vs rate.]{Age vs rate.} \copyright [2018] IEEE. Reprinted, with permission, from \cite{canberk2018}.
  \label{fig:aoi_vs_f}
\end{subfigure}%
\caption{Emulation topology and resulting AoI behavior}.
\label{fig:results_emulator}
\end{figure}

The topology in the emulation study is described in Fig. \ref{core:netwtopo}. The setup ran the CORE 5.0 emulator on the Fedora 27 Linux distribution on a PC with Intel® Core™ i7-7700HQ CPU @ 2.80GHz × 8, 16 GB RAM and 1 TB 7200 rpm HDD. The test setup involves three virtual nodes: server, client and time\_sync. The time\_sync node was for synchronizing the clocks at the client and the server, for correct computation of the age at the server. 

The client generated time-stamped samples at different sampling rates, sending them to the server. Upon receiving each sample, the server computed the status age. Information was exchanged between the server and the client via a link with bandwidth limited to 130 Kbps, marked by the thick line in Fig. \ref{core:netwtopo}, by configuring routers accordingly. All links other than this particular one had unlimited bandwidth, allowing instantaneous packet transmission. None of the links had path delay. Therefore, delay occurred only due to only transport layer queuing. Each iteration of the experiment started with empty buffers, and the sampling rates were gradually increased, obtaining a sample path for AoI vs rate. Four such sample paths are shown in Fig. \ref{fig:aoi_vs_f}.  All iterations exhibit a sharp U-shape behaviour of age with sampling rate. The starting age and the shape of the curve do not appear to deviate much among iterations, and the AoI nearly reaches zero at moderate data rates, owing to the absence of path delay in this emulation. 

\subsection{AoI Measurements on a Real-world Testbed}
\label{Real-world-testbed}

A real-world AoI measurement experiment was reported in \cite{canberk2018} which evaluated age in TCP/IP Connections through 2G, 3G and LTE, WiFi, Ethernet. The study took a two-step approach: first, the server and client were time-synchronized; then, samples were generated at the client at a variable and increasing rate. The client was a computer located \ang{39;53;29.6}N \ang{32;46;56.6}E. The Internet connection for the client was provided alternatively through WiFi, Ethernet, or a cellular connection. The cellular connection was established on the service provider TURKCELL via USB tethering to a cellular phone.  The relatively high bandwidth of the USB 2.0 connection supporting 53 MBps \cite{usb-bandwidth}, is more than capable of handling  data rate resulting from the peak sampling rate. Hence, the effect of the USB buffer on AoI was negligible. 

\begin{figure}
	\centering
	\includegraphics[width=\linewidth]{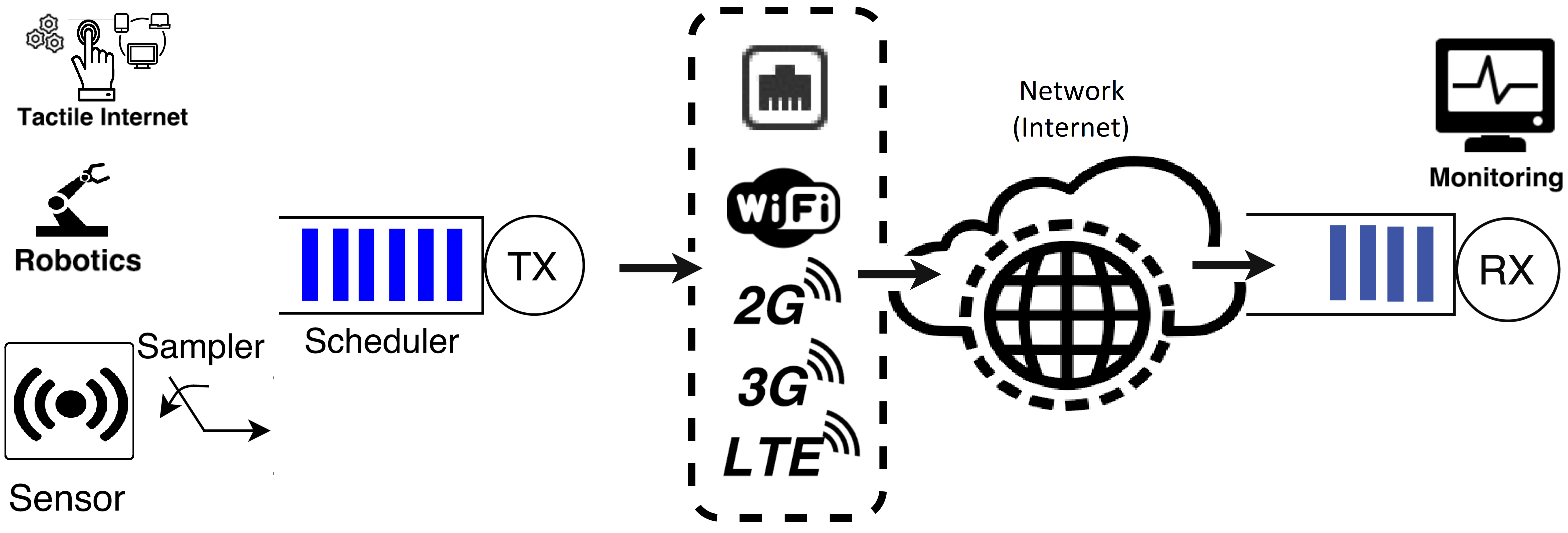}
	\caption[A schematic of the physical testbed.]{A schematic of the physical testbed. \copyright [2018] IEEE. Reprinted, with permission, from \cite{canberk2018}.}
	\label{fig:physical-testbed}
\end{figure}

The topology of the physical testbed, involving a remote server in France, is shown in Fig. \ref{fig:physical-testbed}. The Speed Test application by Ookla was used to measure the upload speeds for each networking scenario: \LTEUploadSpeed for LTE, \ThreeGUploadSpeed for 3G, \TwoGUploadSpeed for 2G, \WiFiUploadSpeed for WiFi and \EthernetUploadSpeed for Ethernet. Each data sample was approximately 35 kB. 

In contrast to the emulation scenario, a time\_sync node is not present in this experimental study. The sender and receiver were in distant geographical locations. Synchronization was performed via prior signaling as described in Section \ref{synchronization}. More precisely, continual requests were sent from the client to the server. The server sent its own timestamp in response to each request. The client recorded the time at the beginning of each request, and then again upon receiving the response to the same request. The round trip delay was computed based on this data. Assuming that the delay was symmetric in both directions in the link, the present clock of the server was predicted. Comparing the predicted clock of the server and its own clock, the client calculated an offset value to synchronize clocks. To reduce offset computation errors, offsets of a large number of independent runs were averaged. A reliable Ethernet connection to a local area network, which had lower latency than the cellular connection, was used as a benchmark for synchronization.

The queues at the transport layer operated on a FIFO basis: the-head of-line packet in the queue was popped and served when the transport channel became available. The sampling rate determined how frequently the new data was pushed into the queue.

\begin{figure}
\begin{subfigure}{.5\textwidth}
  \centering
  \includegraphics[width=1\linewidth]{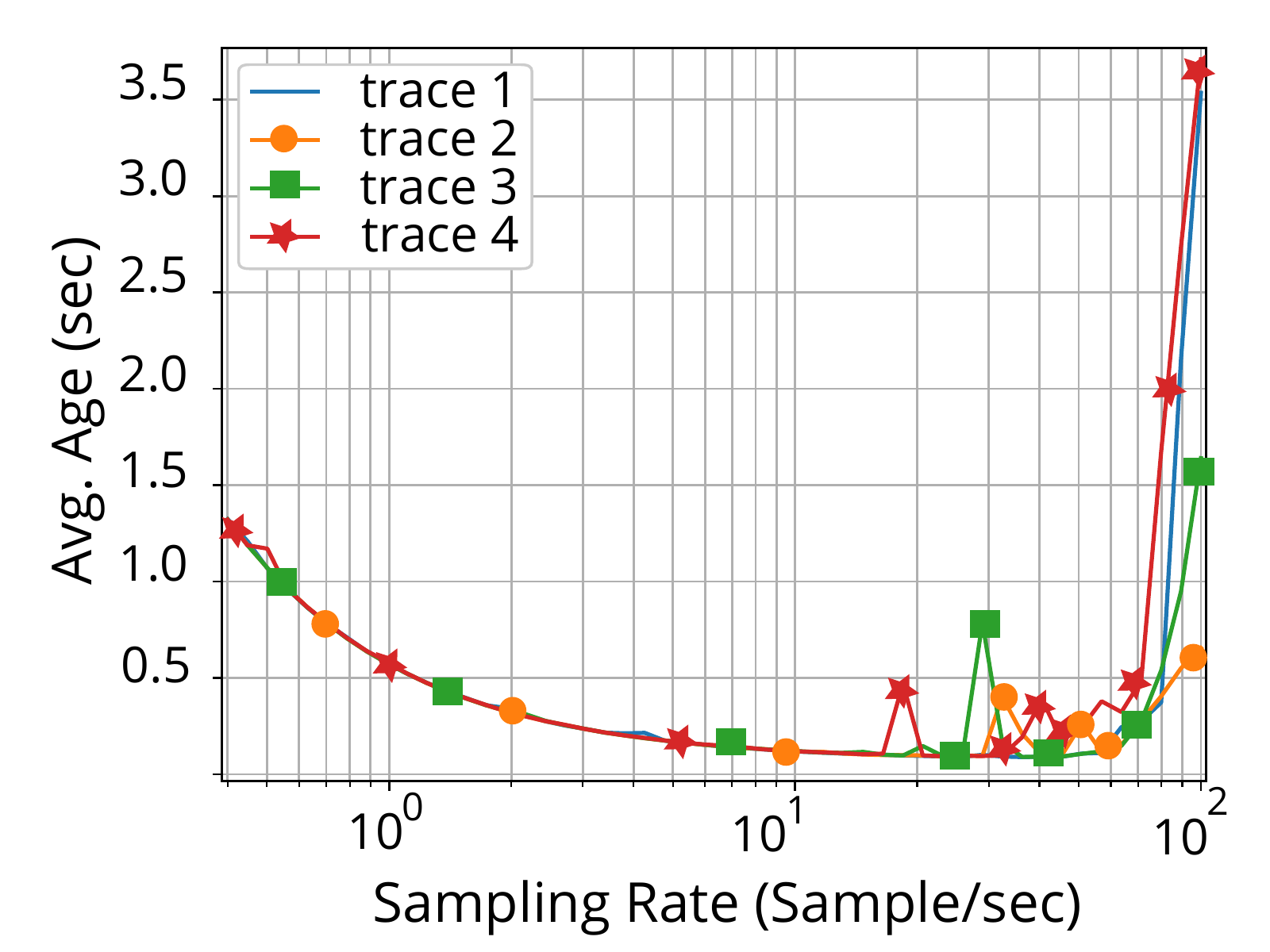}
  \caption{WiFi First-hop}
  \label{real-wifi}
\end{subfigure}%
\begin{subfigure}{.5\textwidth}
  \centering
  \includegraphics[width=1\linewidth]{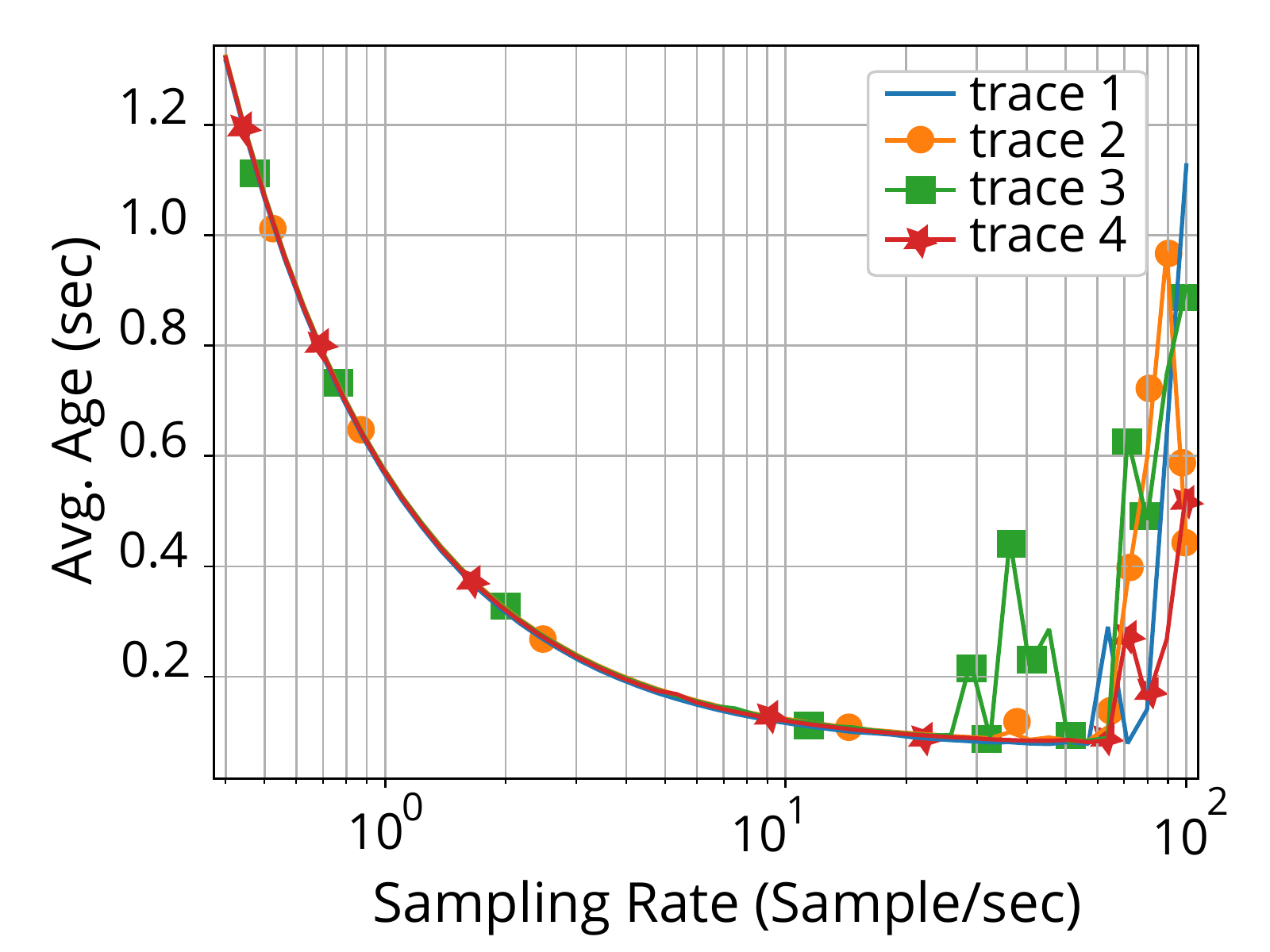}
  \caption{Ethernet First-hop}
  \label{real-ethernet}
\end{subfigure}
\caption[AoI measurements of connections over different access networks.]{AoI measurements of connections over different access networks. \copyright [2018] IEEE. Reprinted, with permission, from  \cite{canberk2018}.}
\label{fig:figAOIwifiEthernet}
\end{figure}

\begin{figure}
\begin{subfigure}{12.9pc}
  \centering
  \includegraphics[width=12.5pc]{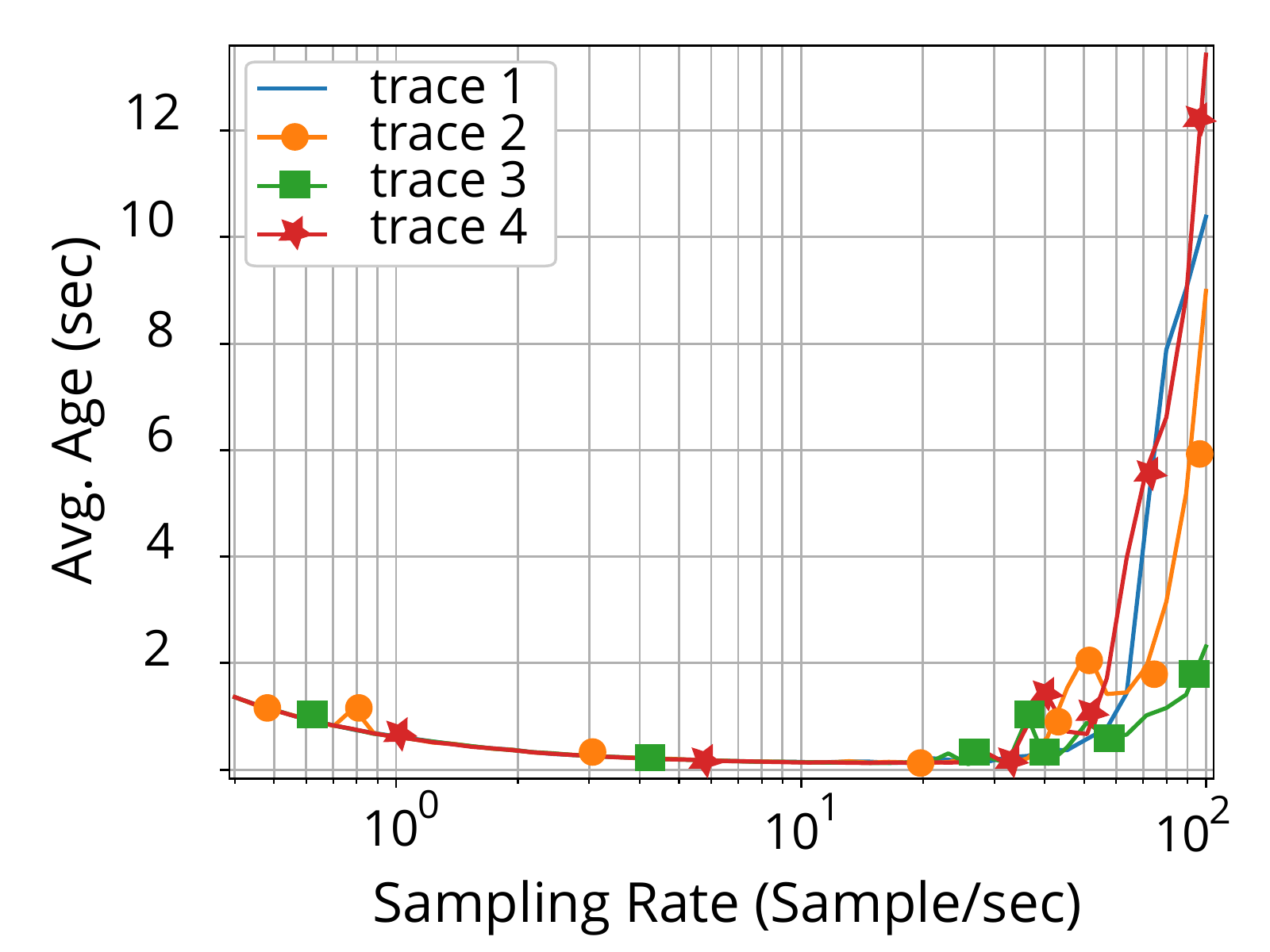}
  \caption{Over LTE Network}
  \label{real-lte}
\end{subfigure}%
\begin{subfigure}{12.9pc}
  \centering
  \includegraphics[width=12.5pc]{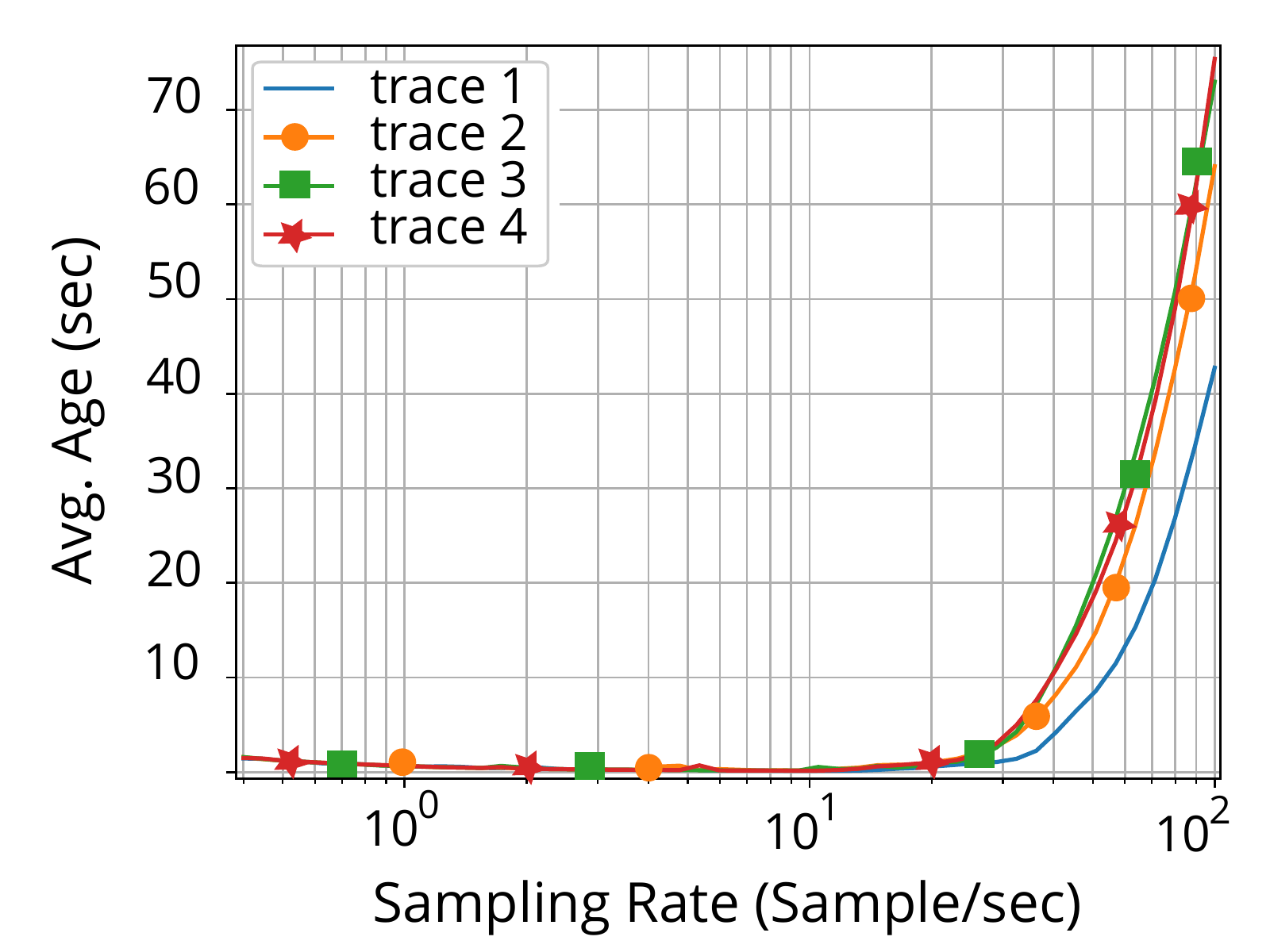}
  \caption{Over 3G Network}
  \label{real-3g}
\end{subfigure}
\begin{subfigure}{12.9pc}
  \centering
  \includegraphics[width=12.5pc]{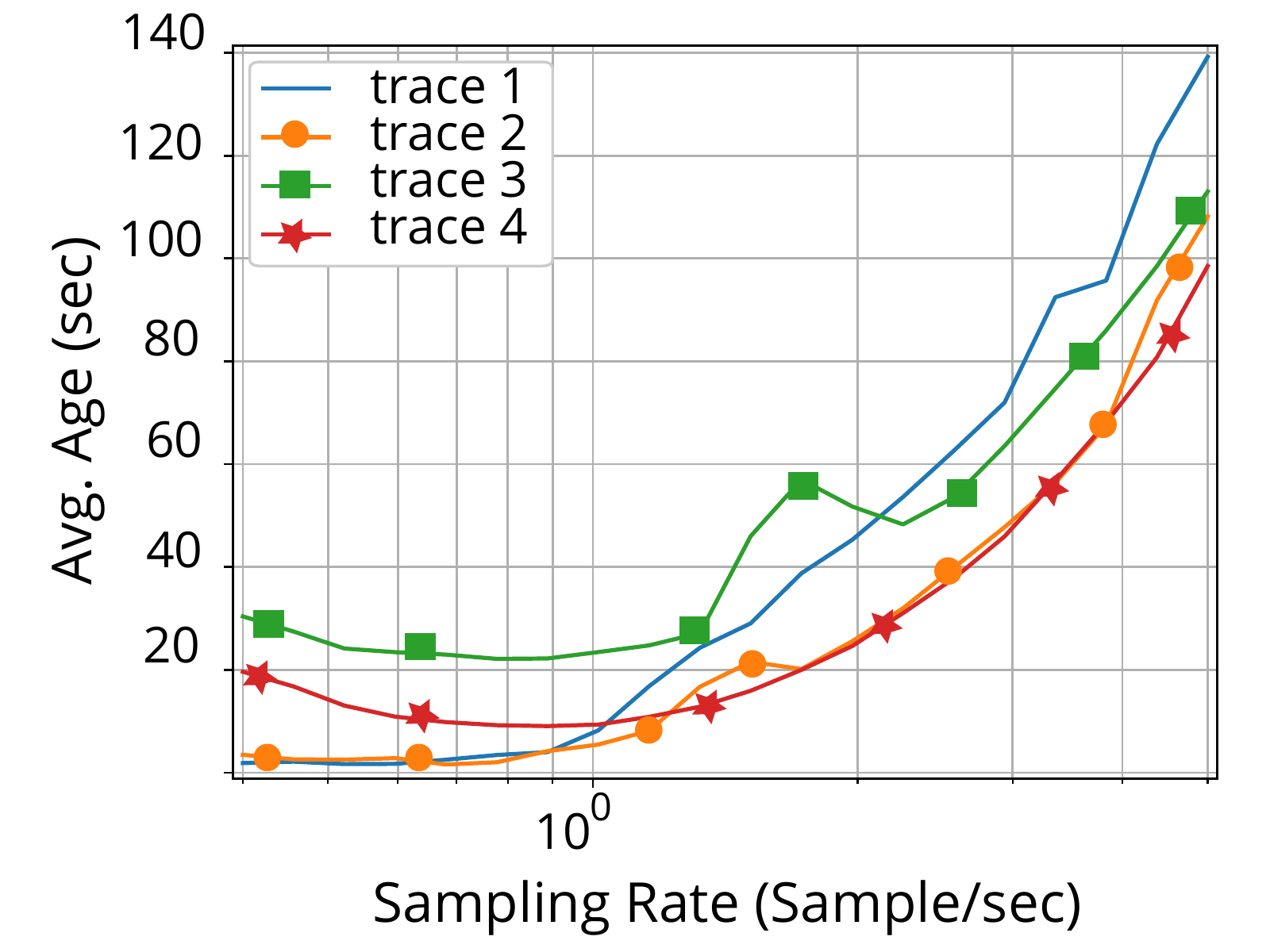}
  \caption{Over 2G Network}
  \label{real-2g}
\end{subfigure}
\caption[AoI measurements on Mobile Networks.]{AoI measurements on Mobile Networks. \copyright [2018] IEEE. Reprinted, with permission, from  \cite{canberk2018}.}
\label{fig:figAOILte3G2G}
\end{figure}




In Figs. \ref{real-wifi}-\ref{real-2g}, the results for different first-hop access types are shown. The figures exhibit a non-monotone behavior of age with sampling rate. The curves look similar at lower sampling rates, however, the AoI starts increasing at different values of the sampling rate, due to the variation in the rates supportable in the first hop links for different access types. In Fig. \ref{real-2g}, it is observed that the AoI increases much faster in case of 2G.

\section{AoI comparison of UDP versus TCP-IP}
\label{AoI_UDPTCP}
TCP and UDP are the two most popularly implemented transport layer protocols in IP networks today. The mechanisms that enable TCP variants to transmit data at high rates without loss, such adaptive window size, congestion control, and re-transmissions make this protocol not only computationally more intense, but also somewhat unsuited to real time data. In terms of age, the re-transmission mechanism of TCP is particularly wasteful as stale packets do not contribute to an age reduction. Many real-time applications use RTP over UDP. However, do the currently available protocols used for real-time traffic truly cater to age objectives? This question justifies a careful look at the age performance in UDP traffic. The brief answer is, while RTP and more modern application layer modifications such as those used in cloud gaming, for example \cite{alos2019cloudgaming} offer relatively timely data for streaming applications, they do not immediately cater to an age objective: status update type flows ultimately do not need all the data to be necessarily transmitted. When there is a newer packet in the sender buffer, it may deem all older ones obsolete. For this reason, to cater to age objectives, we believe that application-layer age-aware sampling or LCFS queuing need to be implemented on top of UDP .

As observed in \cite{6195689,2012CISS-KaulYatesGruteser,Tanaka_2018}, the average age initially decreases with increasing throughput in FCFS systems without any strict buffer management or limitation. Yet, once the communication system struggles to serve the high throughput, the queuing delay becomes significant the average age starts to increase, thereby resulting in the U-shaped relation among average AoI and throughput. The FCFS buffers in routers, switches and access points of practical network infrastructures are therefore expected to lead to a similar non-monotone relationship in real life networks.  

The particular shape of this relation between age and throughput is still influenced by the properties of transport layer protocols, CPU capacity, transmit and receive capabilities etc., as will be demonstrated by the experimental results in the IoT setup, that will be discussed in this section. Achieving a good AoI performance with a low power IoT node first requires that the CPU can generate packets at a sufficient rate. Secondly, the transceiver modules at sender and receiver nodes must be capable of timely processing of the generated packets. Lack of sufficient processing power in either of these will cause an age bottleneck. For example, if the CPU is not fast enough, the transmission rate cannot be increased sufficiently to reduce the AoI. Low-power IoT devices are especially prone to running into these bottlenecks. With common transport protocols, such as TCP, which already have a high computational overhead, and even with UDP, IoT devices can thus run into problems in terms of AoI, underlining the need for new approaches.

In this section, experimental results from \cite{9049792}, taken in real life networks, are reviewed. In particular, we first focus on  the age behavior of TCP and UDP flows over the Internet, Then, we review age behavior of same type of flows among IoT nodes, this time on a local IoT setup, with the goal of demonstrating the effect of limited computational performance of IoT devices on AoI. 
In these experiments, RTT information was used to bound the synchronization error as described in Section \ref{RTTestimate}: before taking measurements, the RTT was estimated by sending several packets and receiving the ACKs. As the ACK packets are small, the estimated RTT was used as an upper bound to the transmission time. Using this as an approximation, the synchronization error was estimated. The remaining age bias was bounded by the RTT. Note that, according to (\ref{eq:linear}), even if approximations still contain a constant bias, this does not affect the location of the age-optimal sampling rate operating point, due to the linearity of the penalty function (which in this case is age itself).

\subsection{UDP vs TCP-IP over Multi-hop Network Testbed}
In the testbed, high power desktop PCs were used to send send TCP or UDP packets through regular Internet/IP infrastructure. Three PCs were located at remote locations, within a 450 km range, yielding a network with paths with varying delays. One of the PC's was set up as the receiver node, and the other PCs sent packets to it. The path between one of the transmitting PCs and the receiver contained approximately 7 hops and 6 ms RTT, whereas the other contained approximately 12 hops and had 80 ms RTT. 

The main cause of AoI increase with UDP was observed to be the high packet loss rate. The delays incurred in transport layer queues were observed to be minor.

\begin{figure}[th!]
\begin{subfigure}{\textwidth}
	\centering
	\includegraphics[width=0.9\linewidth]{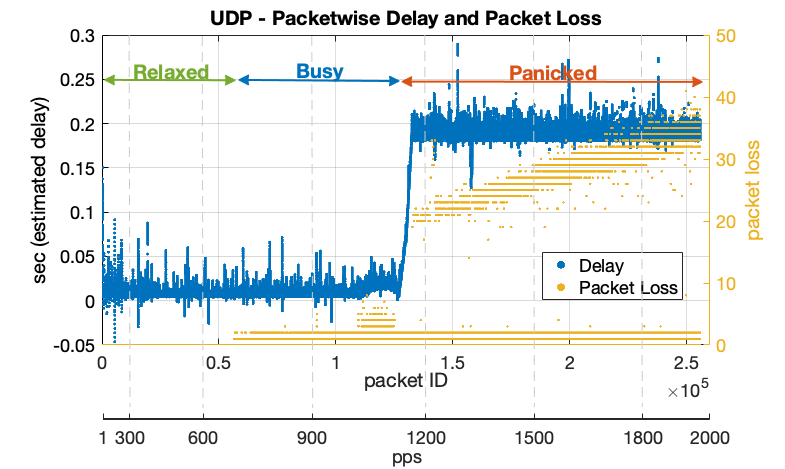}
	\caption{Packetwise delay and packet losses vs rate}
	\label{udpInternet_delay}
\end{subfigure}
\begin{subfigure}{.5\textwidth}
  \centering
  \includegraphics[width=1\linewidth]{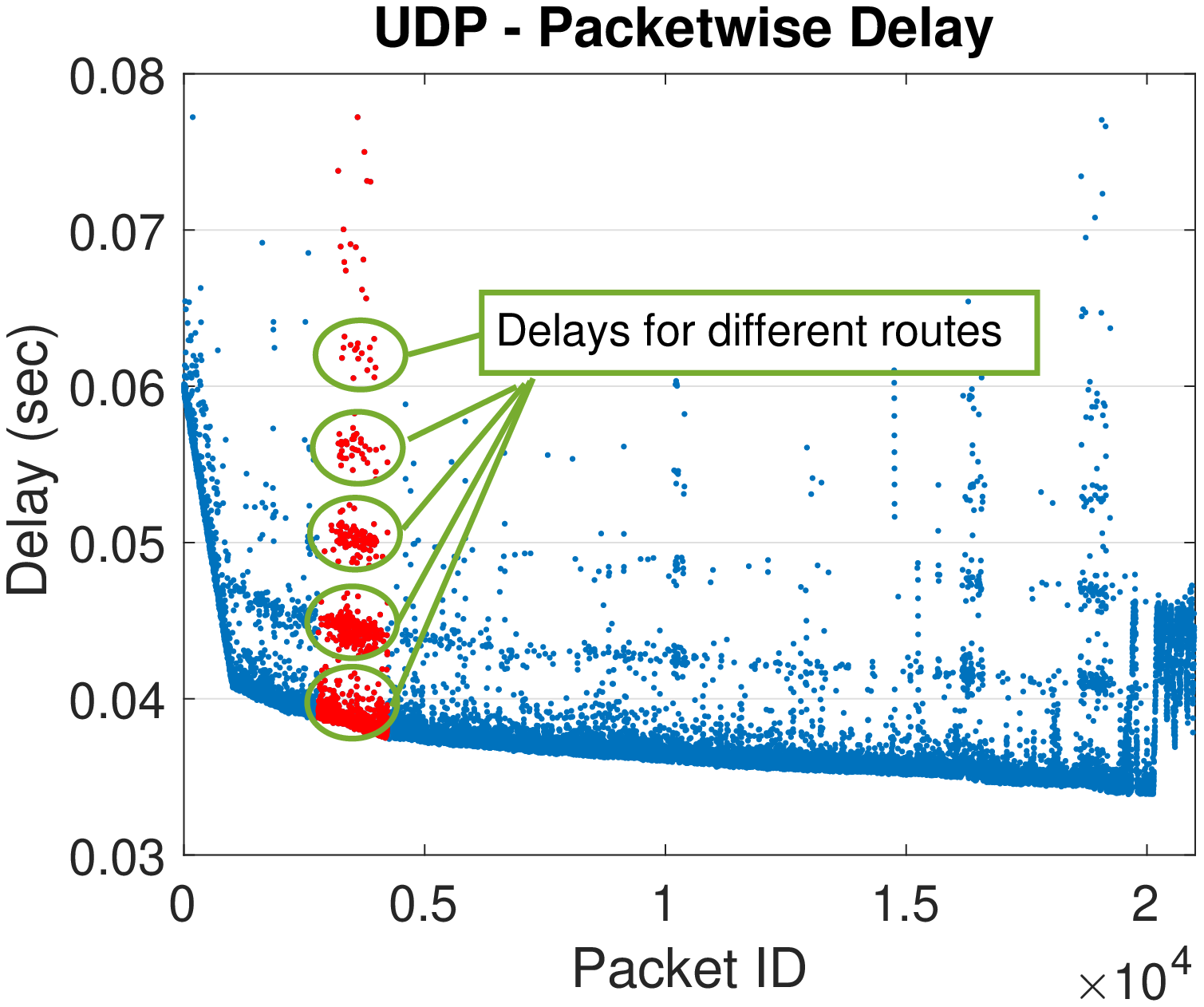}
  \caption{Delay distribution in the relaxed regime}
  \label{fig:udpMultiPath}
\end{subfigure}%
\begin{subfigure}{.5\textwidth}
  \centering
  \includegraphics[width=1\linewidth]{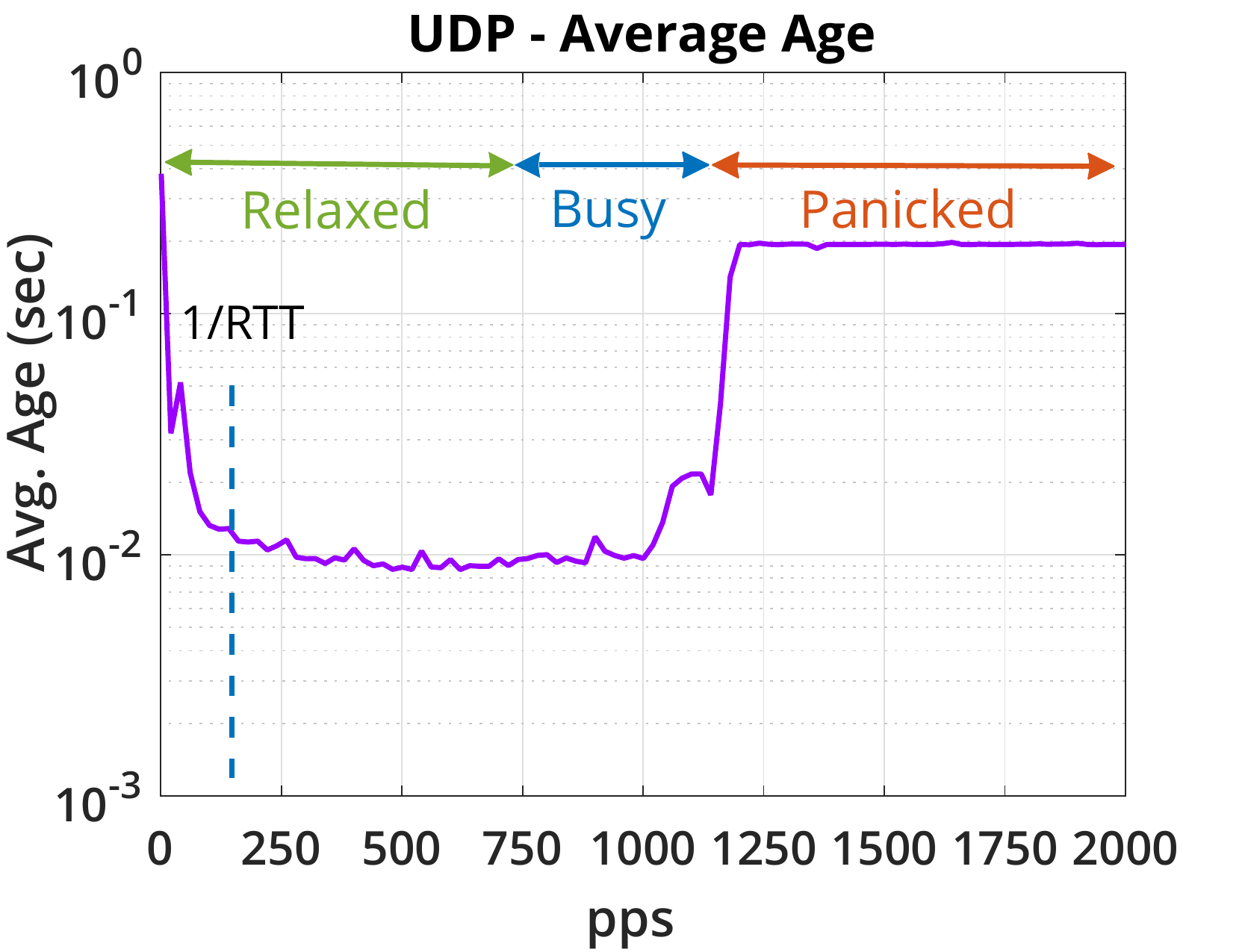}
  \caption{Average age}
  \label{fig:udpage}
\end{subfigure}
\caption[Age, delay and packet loss vs rate for UDP over Internet.]{Age, delay and packet loss vs rate for UDP over Internet. \copyright [2020] IEEE. Reprinted, with permission, from \cite{9049792}.}
\label{fig:figudp}
\end{figure}

\begin{figure}[t]
\begin{subfigure}{0.48\textwidth}
	\centering
	\includegraphics[width=1\linewidth]{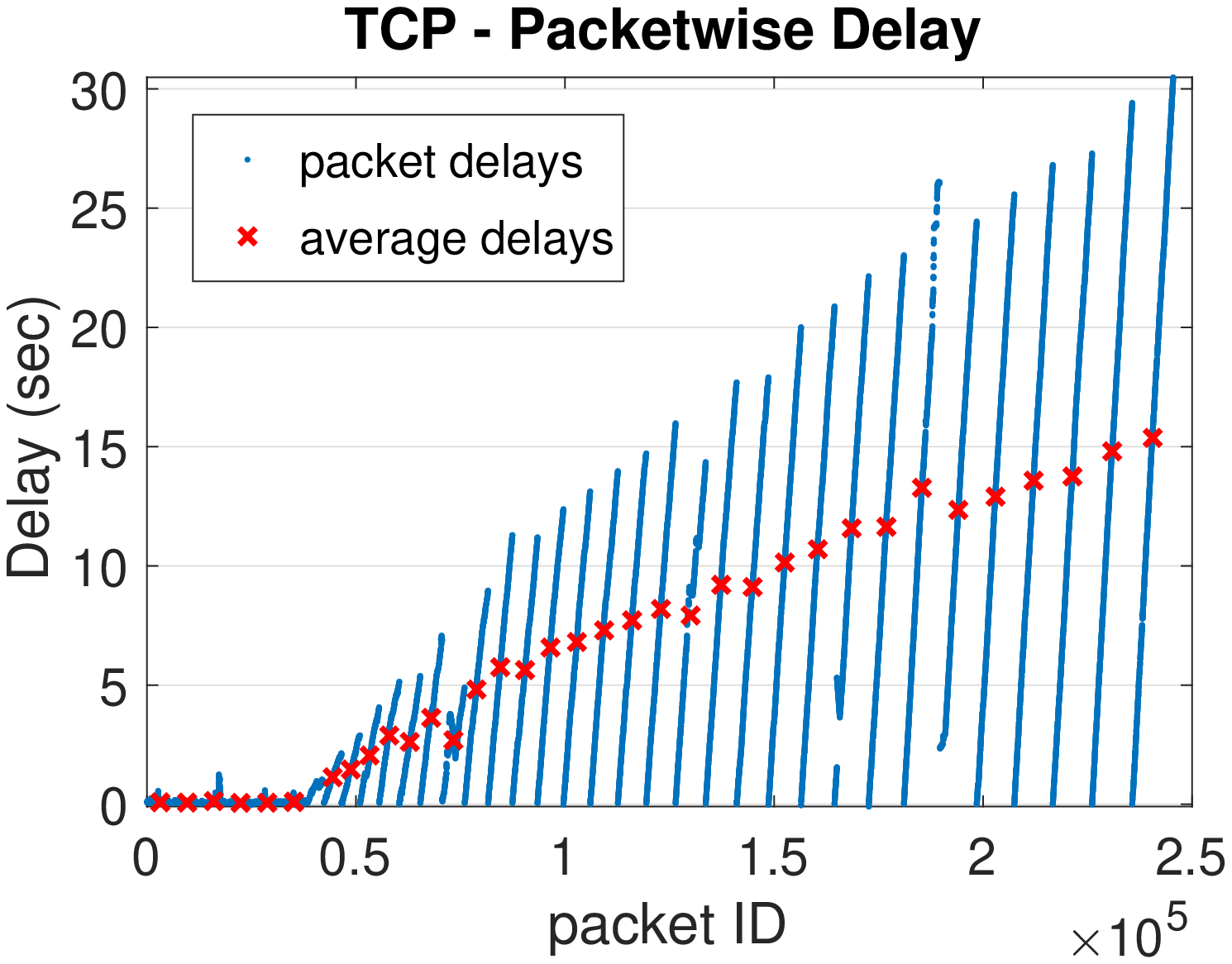}
	\caption{Individual and average packet delays}
	\label{tcpInternet_delay}
\end{subfigure}
\begin{subfigure}{0.48\textwidth}
	\centering
	\includegraphics[width=1\linewidth]{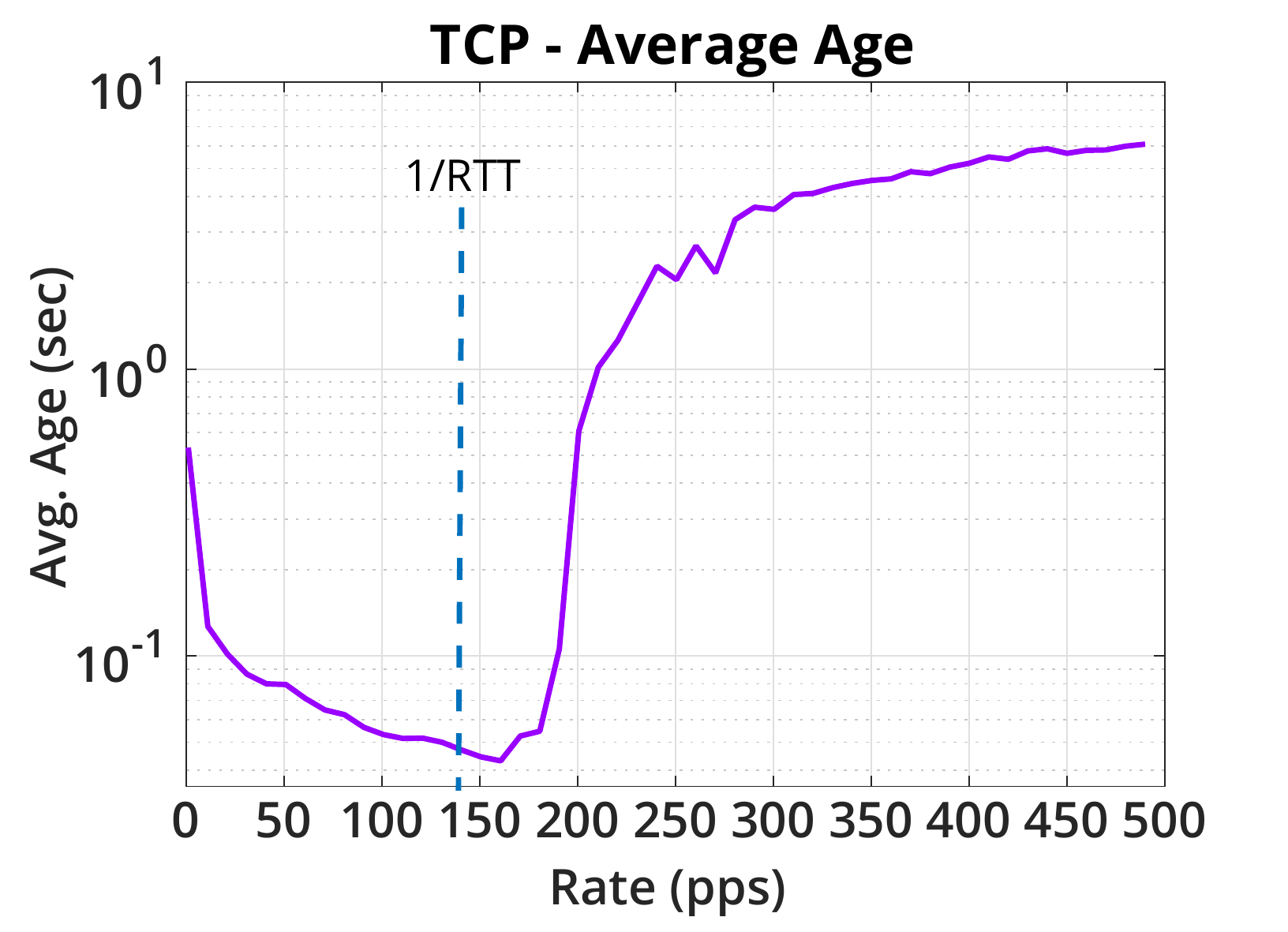}
	\caption{Average age}
	\label{tcpInternet_Age}
\end{subfigure}
	\caption[Age and packetwise delay for TCP over Internet.]{Age and packetwise delay for TCP over Internet. \copyright [2020] IEEE. Reprinted, with permission, from \cite{9049792}.}
\end{figure}

In Figures \ref{udpInternet_delay}-\ref{fig:udpage}, the results of UDP transmission tests on the Internet testbed just described are shown. Figure \ref{udpInternet_delay}, plots delays seen by packets transmitted successfully, as well as the number of lost packets between consecutive successful packets. The horizontal axis represents the packet IDs, and the increasing sample generation rate. In Figure \ref{fig:udpMultiPath}, delays experienced by all packets are plotted, versus increasing rate. The delay distribution suggests the presence of multiple routes, which plays a role in the age remaining constant despite the increasing rate. In Figure \ref{fig:udpage}, the calculated average age values for the same experiments are shown. 

The UDP test results on the Internet testbed reveals three modes of operation, namely, relaxed, busy and panicked. The categorization of these modes is based on the packet loss characteristics. In the relaxed region, number of samples is well below the load that will congest the network. Increasing rate in relaxed mode reduces average age, as more frequent samples provide more timely data.
With increasing load, the network gradually gets congested, causing some nodes to drop packets at random, and enters the busy mode. As seen in Figure \ref{udpInternet_delay}, the start of the busy region is marked by a step up in the number of lost packets from 0 to a moderate number between 1 and 3, and throughout the so called busy region, the number of lost packets generally stays in this range. Despite the occasional packet loss in the busy mode, the delay remains similar to the relaxed mode, as  intermediate nodes can still handle the high traffic without additional queuing delay and the end-to-end delay is mostly due to propagation delay over the links.  

With further increase in rate, we enter the panicked region where both the number of packet losses, and the delay increase. In this mode the queuing delays at intermediate nodes become significant. Note that, due to the lack of a re-transmission mechanism for failed packets in UDP, the packet delays do not monotonically increase with rate, they saturate at a level that depends on the maximum queuing delay. Hence, with increasing rate, the age also remains constant.

Unlike UDP, in TCP, failed packets are re-transmitted, resulting in a steady increase in AoI with increased transmission rates, see Figure \ref{tcpInternet_Age}.   The resulting experimental results, therefore, show a similarity to the U-shaped age-throughput graph of FCFS queues.

Additionally, according to the experiments, TCP flows often observe a sharp increase in  average AoI as rate exceeds a certain value, due to congestion. On the other hand, with UDP, we do not observe a similar sharp age blow up with increased rate.

\subsection{UDP vs TCP-IP over IoT Testbed}
\label{sec:udp_vs_tcp}
The second testbed in \cite{9049792} was a Wi-Fi network, comprising two NodeMCU ESP32 with Xtensa® LX6 (600 MIPS) IoT devices and a Wi-fi router operating in 802.11n mode. The two IoT devices were configured to act as a transmitter and receiver respecively, sending TCP/UDP packets through the central router node. 

Due to the inability of the nodes to run standalone operating systems, the TCP/UDP operations were carried out by Lightweight IP-Stack (LWIP), an open-source software used by different IoT devices.

\begin{figure}[t]
	\centering
	\includegraphics[width=1\linewidth]{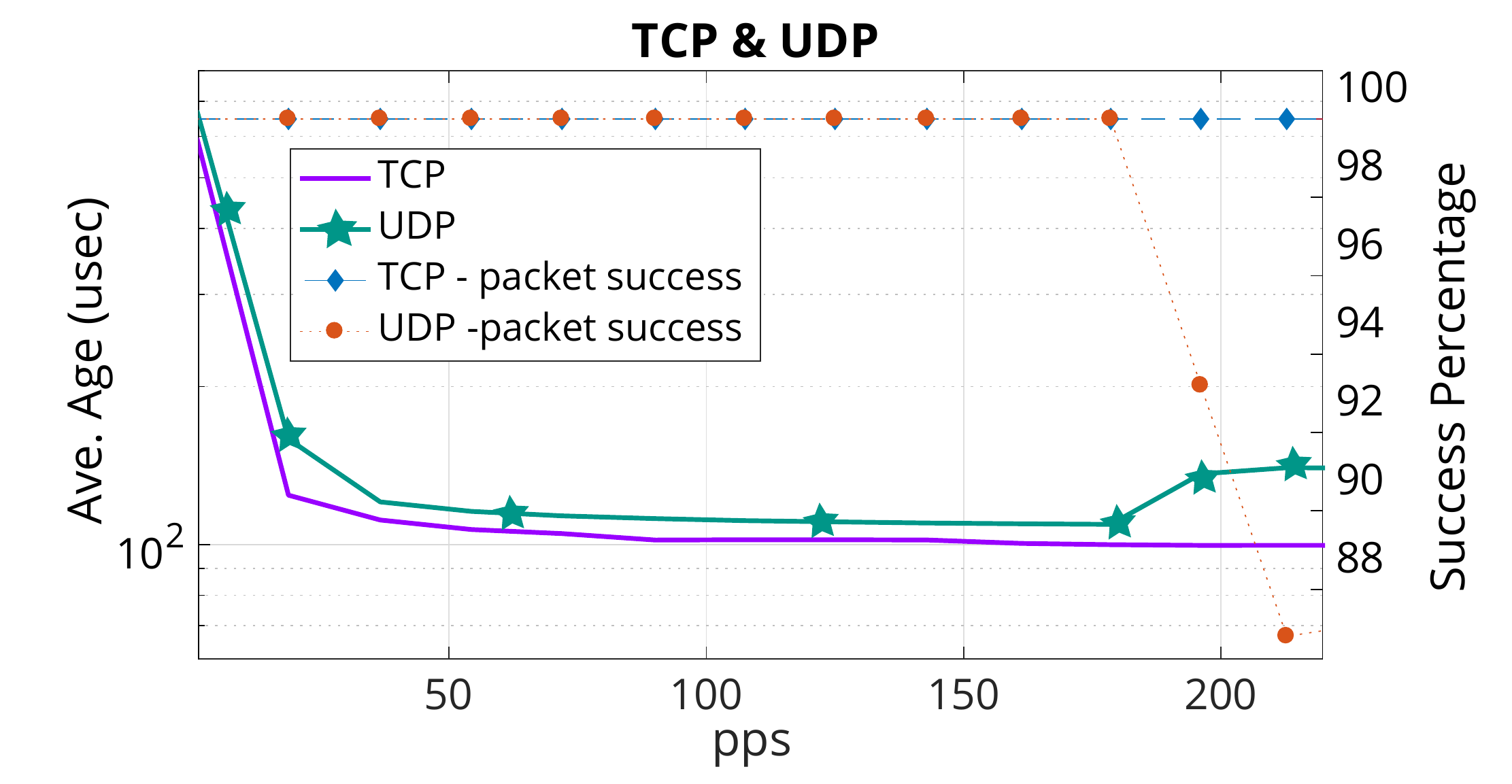}
	\caption[Packet loss and average age TCP and UDP measured over a local Wi-Fi Network, using IoT devices.]{Packet loss and average age TCP and UDP measured over a local Wi-Fi Network, using IoT devices. \copyright [2020] IEEE. Reprinted, with permission, from \cite{9049792}.}
	\label{udptcpIoT_age}
\end{figure}

The results reported in \cite{9049792} indicate an interesting AoI bottleneck for IoT devices: memory and computational power limitations of these devices, rather than any communication or network constraints, can be the dominant cause of aging. The peak packet generation rate for a simple IoT node (e.g., the Wi-Fi module integrated to a ESP32 has higher rate than the processing unit on the device) will be significantly lower than what is possible with a regular PC. Hence the device is able to generate packets at a much lower rate than its transceiver is able to send. Moreover, transceiver buffers on those devices are too small to observe the non-monotone age behavior expected from an infinite-buffer FCFS queuing system. Hence, it is not surprising that Figure \ref{udptcpIoT_age} does not indicate a U-shaped AoI behaviour for either UDP or TCP.  However, with more powerful devices as those typically employed in a LoRa system, non-monotone age behavior can still be observed due to bandwidth bottlenecks.

As a side note, in many IoT applications, as the nodes are typically small, battery-operated devices, energy is a limiting factor that influences the fidelity of operation. Energy limitations have a direct effect on transmission rate and duty cycle, \cite{hakanEurasip2013, energies2014, WirelessEnergySim,FatihGokselLetter,Ceran_2019}, as well as processing capability. We have seen in this section that in simple IoT devices, the processing power tends to be the main limiting factor in age performance, which previous theoretical work has tried to address (e.g., \cite{TanVolkanJCN2019}). 

\section{Application of machine learning methods to AoI optimization in real-world networks}
\label{AoI_MachineLearning}
AoI optimization methods in many previous studies (\eg, \cite{KaulYatesGruteser-Infocom2012,KaulYatesmultiaccess2017,KamKompellaEphremides2014ISIT,YinMultiuserAoI2018, TanTWC2017, TanOnurElifWCNC2018}) relied on knowledge of the statistics of underlying processes (\eg, network delays, energy arrivals). This reduces the applicability of these approaches to practical settings where these statistics are unknown and result from fairly complicated interactions, deeming the application of machine learning methods promising for AoI optimization in real-life networks. To the best of our knowledge, \cite{ceran, EgemenRL} were the first works that utilized machine learning (specifically, reinforcement learning) for minimizing AoI. \cite{ceran} formulated the AoI minimization problem on a point-to-point link through the application of HARQ, as a constrained Markov Decision Process and solved it by using Value Iteration and SARSA algorithms, on discrete state space (ages defined as integers). More recently,  \cite{elgabli2019reinforcement,abd2019deep,ma2020deep,hatami2020age,ferdowsi2020neural} considered the application of machine learning methods to age optimization in various network settings. 

In this section we review \cite{EgemenRL}, which is focused on a real life implementation of a deep RL based algorithm for sample rate optimization in a continuous state-space to control the arrival rate of a flow of packets over the Internet. This is similar to the rate control model in earlier TCP or UDP related works reviewed in previous sections. 

Modeling a flow of packets generated by an application running between a server-client pair over the Internet, \cite{EgemenRL} introduced a deep reinforcement learning-based approach that can learn to control the rate of generation of updates to minimize the AoI with no prior assumptions about network topology. After evaluating the learning model on an emulated network, it was argued that the method can be scaled up to larger and more realistic networks with unknown delay distribution.

Let us briefly discuss the formulation in \cite{EgemenRL}. The sender generates update packets at $t_{1}$, $t_{1}$, $\cdots$, $t_{n}$, these are received at times $U(t_{1}), U(t_{2}), \cdots, U(t_{n})$ (packets can change order in the network.) The AoI optimization problem was modeled as a Markov Decision Process (MDP) with unknown transition probabilities:

\begin{itemize}
\item The state at time $t$ is the age at time $t$:  $s_t = \Delta_t$,
\item There action space at any time $t$ contains two possible actions (pause or resume): $a_t = \{p, r\}$.
\end{itemize}

Accordingly, at any time, the next state depends on the current state and the chosen action through the transition probabilities $p(s_{t+1} | s_t, a_t)$. To learn the transition probabilities, RL was utilized. The goal of the RL is to find the action that maximizes the expected cumulative reward $r(s, a)$ over the trajectory distribution $p_\theta(\tau)$. The trajectory was defined as the state-action pair $\tau = (s, a)$. The AoI minimization objective was formulated with a cost function $c(s_t, a_t) = -r(s_t, a_t) = -\Delta_t$, thus the objective of the formulation in \cite{EgemenRL} is to solve (\ref{eq:aoi_goal}).

\begin{equation}
\theta^* =  \arg\min_\theta \Big(\limsup_{T \to \infty} ~ \frac{1}{T} ~  \mathbb{E}_{(s_t, a_t) \sim p_\theta(s_t, a_t)}\Big[ \Sigma_{t=1}^T \Delta_t \Big]\Big) \label{eq:aoi_goal}
\end{equation}

To solve  (\ref{eq:aoi_goal}), the well known deep Q network (DQN) algorithm, introduced in \cite{dqnnetwork}, was employed. The goal of this learning algorithm was to estimate the expected reward (Q-Value) of possible actions in each state, using neural networks. As it is a cost minimization rather than maximization of a reward, the algorithm was modified such that it minimizes the introduced cost function $c(s_t, a_t) = 1 - \exp\{-\Delta_t\}$.

The algorithm in \cite{dqnnetwork} employed a neural network to map states to the Q-value of each possible action. The Q-values represented the vector of expected costs when each of the actions $a$ was taken in state $s$, and $\arg\min_a ~ \hat{Q}_\phi(s)$ represented the action, at each at state $s$, that minimized the expected cost $\mathbb{E}[c(s, a)]$. Rather than taking the action with the minimum Q-value at each step with probability 1, the agent was forced to explore by utilizing the $\epsilon_{greedy}$ approach \cite{sutton1998reinforcement}, where a random action is taken with probability $\epsilon$. Stabilizing approaches such as experience replay \cite{experience_replay} and Double DQN \cite{vanhasselt2015deep} were also employed.

The DQN algorithm described above was tested in CORE, an open-source network emulator, using the topology in Fig. \ref{core:netwtopo}. In this topology, the path delay on the direct link between routers A and B was selected from a truncated Gaussian distribution with adjustable mean and variance, and the remaining links have zero delay. The routing is planned such that there is only one path between the server and the client, and that involves {routerA} and {routerB}. The  algorithm runs on both the {server} and the {client}. Synchronization between the server and client is performed using a {synchronizer} node. The accuracy of this synchronization is guaranteed as the links between the synchronizer, the {server} and the {client} have no delay. The client assumes one of three modes: (1) generating time stamped samples at a fixed rate, (2) sending samples to the server, and (3) responding to pause and resume commands from the server. The client stops generating new samples when it receives a pause command, and continues upon receiving a resume command.

The case of a link with unlimited bandwidth was tested first. Here, due to the absence of queuing, path delay (1 $\pm$ 0.5s) was the sole cause of aging. Therefore, the pause command can only cause an increase in age this case. Having computed age from the time stamps of the received samples, the age is given to a deep RL algorithm, which uses the age value to decide on a command to reduce the age. This really acted as a test of the RL algorithm: the optimal decision is always \enquote{resume} in this case, as it is possible to transmit data at any sampling rate without loading the network buffer.

\begin{figure}[t]
\begin{subfigure}{0.48\textwidth}
	\centering
	\includegraphics[width=1\linewidth]{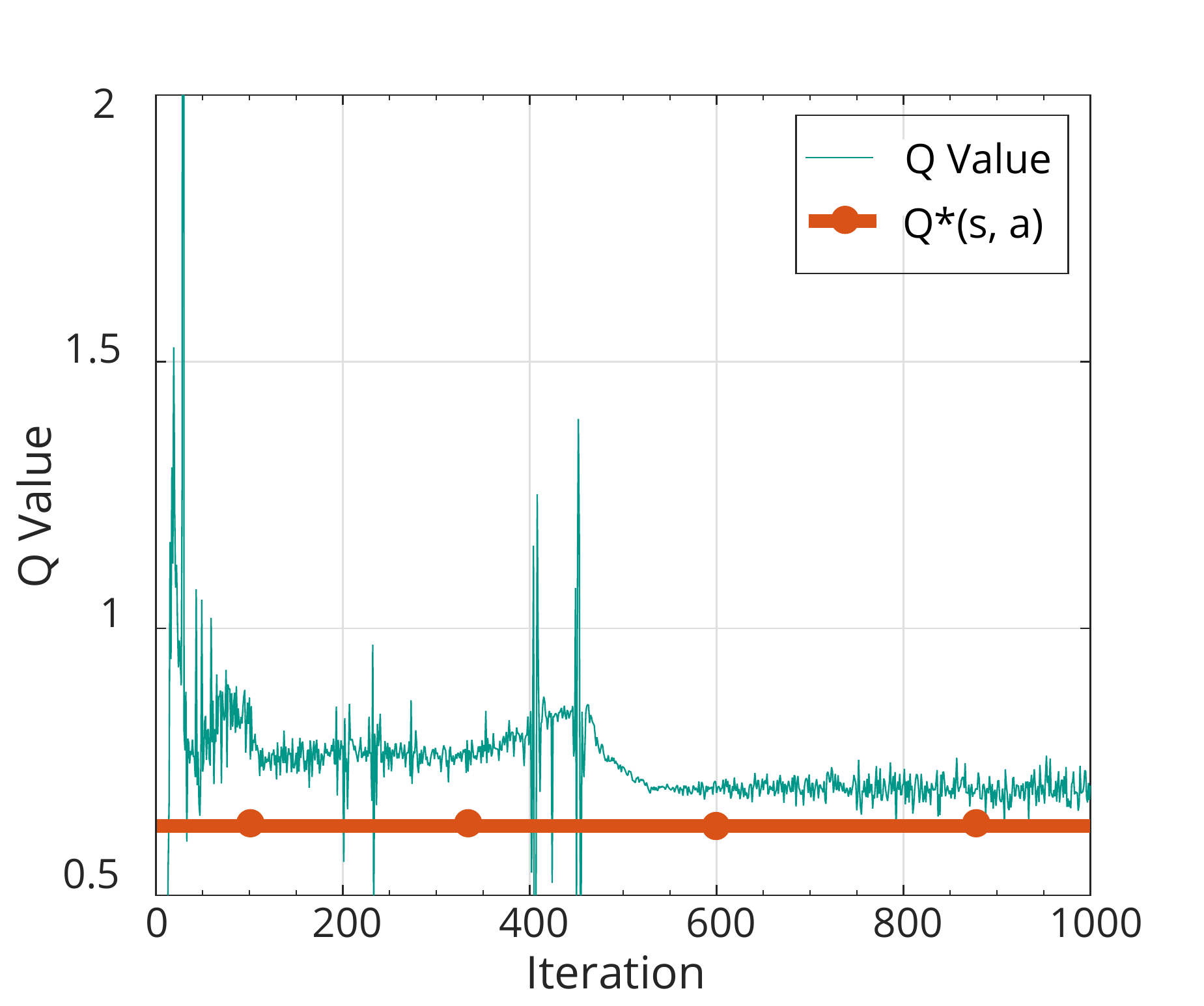}
	\caption{Q-Value per 10 iterations   \\\hspace{\textwidth}   (unlimited bandwith)}
	\label{fig:q_value}
\end{subfigure}
\begin{subfigure}{0.48\textwidth}
	\centering
	\includegraphics[width=1\linewidth]{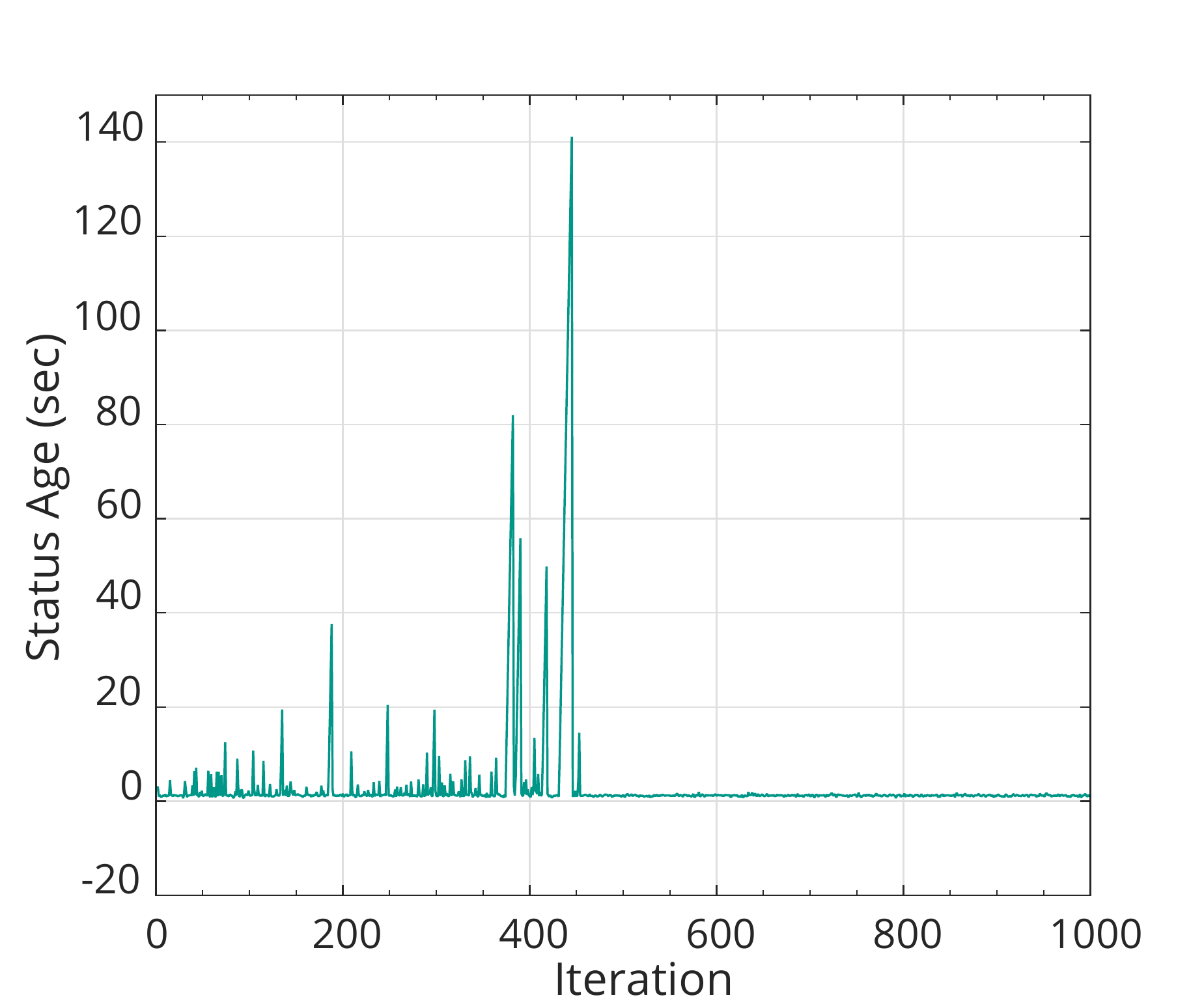}
	\caption{Status age per 10 iterations   \\\hspace{\textwidth}   (limited bandwith)}
	\label{fig:age}
\end{subfigure}
	\caption[Q-Value and Status age  vs. Iteration.]{Q-Value and Status age  vs. Iteration. \copyright [2018] IEEE. Reprinted, with permission, from \cite{EgemenRL}.}
\end{figure}

A double DQN was ran with two hidden layers of 24 units and experience replay, initially trained for 10,000 iterations. The main network was updated each iteration, whereas the target layer was updated every 100 iterations. The Q-Value for the \textit{resume} action is expected to approach  $ Q(s, a) \approx 1 - \exp\{-\Delta_{min\_age}\}$ according to the  boundary conditions applied on the Bellman equation:
\begin{equation}
Q(s, a) = c(s, a) + \gamma \min_{a'} Q(s', a'); \label{eq:bellman}
\end{equation}
At the terminal state, $Q^*(s, a) = c(s, a) = 1 - \exp\{-\Delta_{min\_age}\}$. As the delay is fixed at $1$ s, the Q-Value for the resume action is expected to converge down to $Q(s, a) = 1 - \exp\{-1\} \approx 0.63 $, which happened after approximately 5000 iterations. Fig. \ref{fig:q_value} plots the progress of the instantaneous Q value, as well as its optimal setting. The progress of the status age versus is plotted in Fig. \ref{fig:age}.

Based on the experimental results summarized above, and the Universal Approximation Theorem \cite{uat}, the authors argued that the approach can be scaled up to larger networks with general delay distributions.

\section{Application layer age control mechanisms over UDP}
\label{sec:ACP}
Age Control Protocol \cite{Shreedhar2019} is a transmission control policy designed to work on top of UDP to reduce the Age of Information (AoI) in dynamic networks. The protocol consists of two phases: the initialization phase, and the epochs phase. In the initialization phase, the source (sender) sends a certain number of packets to the monitor (receiver) and calculates RTT. Next, the protocol enters a phase divided into certain periods called epochs. In each epoch, ACP computes the average AoI, an exponentially weighted moving average RTT, exponentially weighted moving average inter-ACK arrival times, and average backlog, i.e., average of the number of packages sent to the receiver but not yet received. At the end of each epoch, the sender's output rate changes according to change in the backlog and average AoI. 

There are three different actions that can be chosen at the end of each epoch: additive increase (INC), additive decrease (DEC), and multiplicative decrease (MDEC). If “INC” or “DEC” actions are used, the protocol tries to increase or decrease the average backlog by a step size $\kappa$. This step size parameter is an important determinant of the resulting age dynamics.

It has been shown that over an inter-continental connection, ACP can reduce the median age by about $33\%$ when compared with \emph{Lazy} \cite{Shreedhar2019}, a simple policy to always keep the total number of packets in flight around $1$ (the delay-bandwidth product \cite{KLEINROCK2018142}) by setting the rate of packet transmission to 1/RTT. The logic of Lazy follows the observation made by Kleinrock in \cite{KLEINROCK2018142} to keep the pipe busy, but not too full. Having sufficiently high load to preventing idling, while still low enough to keep the queues stable, is a classic rule of thumb in finding a throughput-optimal operating point in networks. However, doing this control more finely, on the \enquote{knife-edge} is especially important with respect to the age metric, as any queuing causes unnecessary aging. Keeping the number of packets in flight around $1$ represent this knife-edge: at any time, the network is kept busy transmitting a packet, while there is no packet waiting in queue. It is worth recalling that at the age-optimal operating point of the $M/M/1$ queue, the average number of packets in the system is $\approx 1.13$.

In \cite{UmutReport2021}, the ACP protocol was implemented on IoT devices, specifically two ESP32 nodes controlled via separate Aurdino devices, and connected through a WiFi Hotspot. The results of the implementation suggested that ACP may run into unnecessary oscillations in a small and relatively static network, and that \emph{Lazy} may be preferable when network conditions are more or less static.

\section{Link layer experimentation of age-aware wireless access schedulers}
\label{wireless_access_scheduler}

Building on theoretical results on multiaccess and broadcast scheduling as in \cite{Kadota_scheduling, Kadota2018ACM}, testbed implementations have been reported in \cite{kadota2020wifresh,KeremReport2021}. WiFresh RT \cite{kadota2020wifresh} proposes a novel Wi-Fi uplink implementation using hardware-level programming of schedulers in a network of FPGA-based Software Defined Radios, as well as an application layer scheme for easier adoption of age adaptation, without lower layer protocol stack modifications.  

For each source, WiFresh RT appends time-stamp to each generated packet and stores it in a Last-Come First-Served (LCFS) queue of size 1, that is, keeping only the freshest packet and discarding older packets. WiFresh RT running on the access point (AP) side coordinates the communication in the network.  WiFresh defines two  states for the AP: (1)  waiting  for  a  data  packet, and (2)  transmitting  a  poll  packet.  The WiFresh App runs over UDP protocol and uses standard Wi-Fi. This App contains all elements of WiFresh RT and some additional features. Fragmentation of large information updates, a simple built-in synchronization algorithm, and support for sources that generate multiple types of information are additional features of WiFresh App. 

The experimental results in \cite{kadota2020wifresh} indicate that when congestion in the wireless network increases, the Age-of-Information degrades sharply, leading to outdated information at the destination. WiFresh is shown to mitigate this problem, achieving near optimal information freshness in wireless networks of any size, even when the network is overloaded. WiFresh was compared with UDP over Wi-Fi and Wi-Fi Age Control Protocol (ACP) \cite{Shreedhar2019} on architectures using a Raspberry Pi wireless base station receiving data from $N$ sources. The results show that ACP improves the average age by a factor of four when compared with Wi-Fi UDP, while WiFresh improves the average age by a factor of forty when compared with Wi-Fi UDP. The reason for the superiority of WiFresh over other two architectures is that WiFresh has used a combination of a polling multiple access mechanism, the Max-Weight (MW) policy and, LCFS queues at the Application layer. Unlike ACP, WiFresh and Wi-Fi UDP architectures do not attempt to control the packet generation rate at the sources. Therefore, the when number of sources N increases to the point that the cumulative packet generation rate exceeds the capacity of the network, both Wi-Fi UDP and WiFresh become overloaded. Consequently, number of backlogged packets at the sources grows rapidly. As Wi-Fi UDP uses the FCFS policy, a large backlog in queues leads to high packet delay, resulting in high average age. In contrast, WiFresh scales gracefully with average age increasing linearly with N.

Results of an experimental implementation of a wireless downlink are reported in \cite{KeremReport2021}. On a multiuser testbed consists of multiple software-defined USRP radios, the authors have implemented Round-Robin, Max-Weight, Whittle’s Index and Greedy policies provided in \cite{Kadota2018ACM} using Labview and tested them under different node formations and power limitations. In this downlink implementation, at each time frame, the AP generates a status update and sends it to the receiver determined by the scheduling policy. The transmitter uses QPSK modulation and BCH coding. The  transmitter and receiver nodes have access to a common clock due to being connected to a common computer. A performance comparison of scheduling policies based on the measured age values is reported to indicate that Max-Weight and Whittle’s Index policies outperform Round-Robin and Greedy policies, especially when signal power is strictly limited, and the channels are asymmetric, i.e., observing different channel gains.

\section{Conclusions}
\label{conclusion}

In this chapter, we have provided an overview of implementation studies related to Age of Information as a performance metric and design criterion on communication networks. These studies have each provided a proof-of-concept of how data freshness can be improved by modifying the design of various protocols and algorithms in the networking stack, without loss of performance with respect to traditional metrics such as throughput and delay. They have each addressed certain practical issues that arise in the process of incorporating AoI into the design of network architectures and algorithms, and discussed how to handle those.

We believe, based on the theoretical understanding about AoI having gained maturity within the last decade, and the results of these pioneering implementation studies, that it is worthwhile to introduce age-aware modifications to the network stack. Perhaps it will be easiest for these principles to penetrate application layer mechanisms, and in fact, most of the rate control mechanisms that we have reviewed in this paper have done exactly that. 

In conclusion, we believe we are witnessing the first steps of the concept of AoI entering mainstream network protocol design. This may be followed by more general semantics related metrics some of which may capture more about the information than its timeliness. One has to be careful, though, in finding a balance between the usefulness of semantics-related metrics and their practical applicability in the protocol stack. While semantics-empowered communication has been a holy grail in communication and information theory for many decades, it has not found its way to implementation mainly because of the requirement of cross-layer operation across the protocol stack. For example, optimizing physical layer coding schemes to directly serve the needs of a control system \cite{mitterLQG1998} is potentially more efficient than ignoring the semantics and treating all data equally, however, this would require the transceiver operation to be customized to interact with the application layer, which is difficult from a standardization and industry-wide adoption perspective. 

Age is perhaps the simplest of semantics related metrics as it depends only on time-stamps. While it may be considered simplistic for the same reason, many previous studies have shown that application performance (e.g. location tracking) can either be characterized in terms o a function of age, or even if this is not true, age-based sampling still provides a major step toward optimal performance (e.g.,\cite{SunPolyanskiyUysalTransIT2019}). We have seen in the implementation examples reviewed in this chapter that it is possible to obtain significant age improvements by simple modifications such as  application layer sample generation rate. We are thus convinced that such revision of protocols has great potential, due to age, as a freshness metric, capturing the needs of IoT and other applications often requiring status-update type data, and are constituting an increasingly large portion of Internet traffic.





  



\bibliographystyle{IEEEtran} 


\bibliography{BookAOIParxiv}






\end{document}